\documentclass[fleqn,trsc,nonblindrev]{informs4}

\OneAndAHalfSpacedXI % current default line spacing
\usepackage{endnotes}
\let\footnote=\endnote

\usepackage{graphicx}
\usepackage{multirow}
\usepackage{hhline}

\usepackage[small, margin=1cm]{caption}

\usepackage{appendix}
\usepackage{color}
\definecolor{strcolor}{rgb}{0.6, 0.2, 0.6}
\definecolor{commentcolor}{rgb}{0.3125, 0.5, 0.3125}
\definecolor{keycol}{rgb}{0, 0, 1}

\usepackage[ruled,vlined]{algorithm2e}
\usepackage[noend]{algpseudocode}
\usepackage{bm}

\usepackage{bbm}

\usepackage{listings}
\usepackage{hyperref}
\usepackage{url}

\newcommand {\bea}{\begin{eqnarray}}
	\newcommand {\eea}{\end{eqnarray}}

\def\blot{\quad \mbox{$\vcenter{ \vbox{ \hrule height.4pt
				\hbox{\vrule width.4pt height.9ex \kern.9ex \vrule width.4pt}
				\hrule height.4pt}}$}}

\usepackage{natbib}
\bibpunct[, ]{(}{)}{,}{a}{}{,}%
\def\bibfont{\fontsize{8}{9.5}\selectfont}%
\TheoremsNumberedThrough     % Preferred (Theorem 1, Lemma 1, Theorem 2)
\ECRepeatTheorems

\EquationsNumberedThrough    % Default: (1), (2), ...
\gdef\AQ#1{}
\gdef\CQ#1{}
\begin{document}
	%%%%%%%%%%%%%%%%
	
%	\AIA
% \setcounter{page}{1} %
% \VOL{00}%
% \NO{0}%
% \MONTH{Xxxxx}%
% \YEAR{2017}%
% \FIRSTPAGE{1}%
% \LASTPAGE{16}%
% \FIRSTPAGEAIA{1}%
% \LASTPAGEAIA{16}%
\def\COPYRIGHTHOLDER{INFORMS}%
\def\COPYRIGHTYEAR{2017}%
\def\DOI{\fontsize{7.5}{9.5}\selectfont\sf\bfseries\noindent https://doi.org/10.1287/opre.2017.1714\CQ{Word count = 9740}}
%\def\RECEIVED{November 1, 2016}
%\def\REVISED{June 22, 2017; October 6, 2017}
%\def\ACCEPTED{November 15, 2017}
% \PUBONLINEAIA{}

	\RUNAUTHOR{Wang et~al.} %

	\RUNTITLE{Traffic State Estimation and Uncertainty Quantification at Signalized Intersections}

\TITLE{Traffic State Estimation and Uncertainty Quantification at Signalized Intersections with Low Penetration Rate Vehicle Trajectory Data}


	\ARTICLEAUTHORS{
%		\AUTHOR{Jianzhe Zhen,\textsuperscript{a,*} Dick den
%		Hertog,\textsuperscript{a} Melvyn Sim\textsuperscript{b}} 
%\AFF{$^{a}$Department of Econometrics and Operations Research,
%Tilburg University; $^{b}$NUS Business School, National University of
%Singapore}

\AUTHOR{Xingmin Wang$^\dagger$, Zihao Wang$^\dagger$, Zachary Jerome}
\AFF{Department of Civil and Environmental Engineering, University of Michigan}

\AUTHOR{Henry X. Liu$^*$}
\AFF{Department of Civil and Environmental Engineering, University of Michigan \\
Mcity, University of Michigan }

\AUTHOR{$^*$Corresponding author; $^\dagger$these authors contributed equally.}

% \AUEXTRA{* Corresponding author}

%\AFFmail{{\bf Contact:} j.zhen@tilburguniversity.edu,
%d.denhertog@tilburguniversity.edu,\\			melvynsim@nus.edu.sg}%
}
	 % end of the block
	
% \ARTICLEINFO{\textbf{Received:} November 1, 2016\\ \textbf{Revised:} June 22, 2017; October 6, 2017\\ \textbf{Accepted:} November 15, 2017\\ \textbf{Published Online in Articles in Advance:}}

	\ABSTRACT{This paper studies the traffic state estimation problem at signalized intersections with low penetration rate vehicle trajectory data. While many existing studies have proposed different methods to estimate unknown traffic states and parameters (e.g., penetration rate, queue length) with this data, most of them only provide a point estimation without knowing the uncertainty of these estimated values. It is important to quantify the estimation uncertainty caused by limited available data since it can explicitly inform us whether the available data is sufficient to satisfy the desired estimation accuracy. To fill this gap, we formulate the partially observable system as a hidden Markov model (HMM) based on the recently developed probabilistic time-space (PTS) model. The PTS model is a stochastic traffic flow model that is designed for modeling traffic flow dynamics near signalized intersections. Based on the HMM formulation, a single recursive program is developed for the Bayesian estimation of both traffic states and parameters. As a Bayesian approach, the proposed method provides the distributional estimation outcomes and directly quantifies the estimation uncertainty. We validate the proposed method with simulation studies and showcase its applicability to real-world vehicle trajectory data.

 % Simulation studies validate the efficacy of the proposed method, with a subsequent case study using real-world trajectories. 
 %    We validate the efficacy of the proposed method with simulation results and demonstrate

    % Simulation studies validate the efficacy of the proposed method. Additionally, we showcase its applicability to real-world vehicle trajectory data via a case study.
 }

% The partially observable system is formulated as a hidden Markov model and a single recursive program is utilized to estimate both unknown traffic parameters and real-time traffic states.

%\FUNDING{The research of the first author is supported by NWO Grant 613.001.208. The third author acknowledges the funding support from the Singapore Ministry of Education Social Science Research Thematic Grant MOE2016-SSRTG-059.}

\SUBJECTCLASS{\AQ{Please confirm subject classifications.}}

\AREAOFREVIEW{Traffic state estimation.}

%{\CQ{Kindly provide the keywords.}}

	%%%%%%%%%%%%%%%%%%%%%%%%%%%%%%%%%%%%%%%%%%%%%%%%%%%%%%%%%%%%%%%%%%%%%%
	
	% Samples of sectioning (and labeling) in OPRE
	% NOTE: (1) \section and \subsection do NOT end with a period
	%       (2) \subsubsection and lower need end punctuation
	%       (3) capitalization is as shown (title style).
	%
	%\section{Introduction.}\label{intro} %%1.
	%\subsection{Duality and the Classical EOQ Problem.}\label{class-EOQ} %% 1.1.
	%\subsection{Outline.}\label{outline1} %% 1.2.
	%\subsubsection{Cyclic Schedules for the General Deterministic SMDP.}
	%  \label{cyclic-schedules} %% 1.2.1
	%\section{Problem Description.}\label{problemdescription} %% 2.
	% Text of your paper here
	
\maketitle
		
\section{Introduction}

Traffic state estimation has been an important problem for both research and practice since it directly provides input for traffic control and operations. It refers to the estimation of traffic state representations given different available data sources (e.g., loop detectors, and probe vehicles). 
% For example, traditional Eulerian traffic state representations include the flow, density, and speed of road segments given pre-determined spatial and temporal resolutions. 
During the past decades, researchers have proposed different traffic state estimation methods. Readers can refer to the recent survey papers for a more comprehensive review of both freeway \citep{seo2017traffic,wang2022real} and urban traffic state estimation \citep{guo2019urban,xing2022traffic,maripini2023traffic}. 

This paper studies traffic state estimation at signalized intersections with low penetration rate vehicle trajectory data. Vehicle trajectory data has become increasingly available during the past years from different resources such as navigation apps, ride-hailing services, etc. Without relying on any physical road infrastructure, vehicle trajectory data provides a more scalable, sustainable, and efficient solution for traffic monitoring and evaluation \citep{waddell2020scalable,saldivar2021deriving,wang2023trajectory}. Compared with fixed-location sensors that can only provide observation within the detection range, vehicle trajectory data covers large spatial-temporal spaces, particularly for those roadways with higher traffic volumes. However, the penetration rate of the currently available vehicle trajectory data is usually limited. The sparse and incomplete observation caused by a low penetration rate is the main bottleneck of utilizing vehicle trajectory data for traffic state estimation and other applications. 

Many existing studies have proposed different methods to estimate real-time traffic state or other related parameters (i.e., penetration rate, traffic volumes) by exclusively using vehicle trajectories. 
% While data-driven approaches (i.e., deep learning) have drawn tremendous attention during past years \citep{wang2023machine}, model-based approaches still dominate those applications with low penetration rate vehicle trajectories due to the unsupervised nature: the absence of overall traffic makes it difficult for data-driven methods, at least they need to be combined with certain model-related regularization \citep{di2023physics}. 
Readers can refer to Section 2 of \cite{guo2019urban} for a more complete review of the existing studies, which can be roughly divided into two categories: deterministic and stochastic.  Deterministic methods use a deterministic model and most of them are developed based on shockwave theory \citep{light1955kinematic,richards1956shock}, such as \cite{ban2011real,cheng2012exploratory,sun2013vehicle,ramezani2015queue,li2017real}. However, deterministic methods ignore the randomness of traffic demand and barely leverage prior information that can be obtained from historical data. Stochastic methods, on the other side,  address these issues and tend to be more reliable, particularly under a lower penetration rate. The earliest stochastic methods probably come from a series of works by \cite{comert2009queue, comert2011analytical, comert2013simple}, which focused on the stop locations of observed vehicles at certain snapshots (e.g., the start of the green time of each cycle). These works inspired many later studies which also formulated the problem based on the stop locations of observed vehicles \citep{zhao2019estimation,zhao2019various,zhao2021hidden,zhao2021maximum,wong2019estimation,jia2023uncertainty}. Another notable work is from \cite{zheng2017estimating}, which applied a maximum likelihood estimation (MLE) to estimate the traffic volume. The expectation-maximization (EM) algorithm \citep{neal1998view} was used as the solution algorithm, which turned out to be a good fit for the problem due to the existence of the latent variables. This framework is widely adopted afterward although the specific formulations are different \citep{yao2019sampled,zhao2021maximum}. 

Limited by the sparse and incomplete observation at a low penetration rate, most existing methods mentioned before need to rely on aggregating historical data and assuming that certain traffic parameters (i.e., penetration rate, arrival rate) are stationary within the aggregation period. Most of them used point estimators (including MLE) and cannot explicitly quantify the estimation uncertainties caused by limited available data. In other words, they cannot indicate how much data is sufficient to provide an acceptable estimation result. This is an important issue since if the utilized data is insufficient, the estimation result will become less reliable. As a result, we might need to use a longer aggregation period (either more days of data or a longer TOD period). However, the stationary assumption will become less valid in this case since the traffic is actually time-varying and also changes from day to day. The aggregation period should be as short as possible while ensuring it is sufficient to satisfy the desired estimation accuracy. 

% Regarding this trade-off, the ideal choice is to use as short an aggregation period as possible while ensuring it is sufficient to satisfy the desired estimation accuracy. 

To fill this gap, this paper utilizes the Bayesian approach to provide a distributional estimation for all estimated values. The major advantage of the Bayesian method is that it can explicitly quantify the uncertainty caused by limited available data. Thus, it can properly address the data sufficiency issue by clearly informing us whether the available data is sufficient to provide an accurate estimation. The proposed estimation method is formulated based on a recently developed stochastic traffic flow model called the probabilistic time-space (PTS) model. The PTS model is proposed by \cite{wang2024osaas} and is particularly designed for modeling traffic flow dynamics near signalized intersections. 
% By assuming that all vehicles follow a uniform deterministic Newell's car-following model \citep{newell2002simplified}, the PTS model establishes a bidirectional mapping between the spatial-temporal vehicle trajectories and a simple point-queue process in Newellian coordinates. In this way, a point-queue model can be directly used to represent the spatial-temporal traffic state. 
% It enables us to formulate a Bayesian network that connects unknown traffic states and observed vehicle trajectories. Consequently, different statistical estimation approaches can be utilized. For example, \cite{wang2024osaas} applies the method of moments estimator to estimate the penetration rate by matching the observed average delay and the model-estimated average delay. 
Based on the PTS model, the partially observable system is formulated as a hidden Markov model (HMM). The observable state includes the observed arrival and queue length extracted from the observed trajectory while the hidden state is the overall traffic state including both observed and unobserved vehicles. Besides, there are some unknown traffic parameters governing this HMM, such as the arrival rate and penetration rate. Based on this HMM formulation, the estimation problem is decomposed into two sub-problems: 1) parameter estimation and 2) hidden state estimation. A single recursive algorithm is utilized to find both the posterior of the hidden traffic state and the (marginal) likelihood function given certain traffic parameters. This marginal likelihood function can then be used by different Bayesian estimation approaches, which provide distributional estimation results for these unknown traffic parameters. 

The main contributions and highlights of this paper are listed below:
\begin{enumerate}
     \item We apply Bayesian method to estimate both real-time traffic state and stationary latent traffic parameters at signalized intersections based on the probabilistic time-space (PTS) model; 
    \item Through a Bayesian approach, the proposed method can provide distributional estimation results, which explicitly quantify the uncertainty caused by limited available data.
\end{enumerate}

The rest of the paper is organized as follows. Section \ref{sec:statement} provides a more detailed problem statement as well as the main assumptions utilized in this paper. Section \ref{sec:formulation} introduces the probabilistic model and Section \ref{sec:estimation} introduces the associated solution algorithms. The proposed method is mainly validated through simulation studies in Section \ref{sec:simulation} while a case study with real-world vehicle trajectory data is also available in Section \ref{sec:real-world}. Section \ref{sec:conclusion} concludes this paper at last. 

\section{Problem statement and assumptions}  \label{sec:statement}

\begin{figure}[h!]
    \centering
    \includegraphics[width=0.85\linewidth]{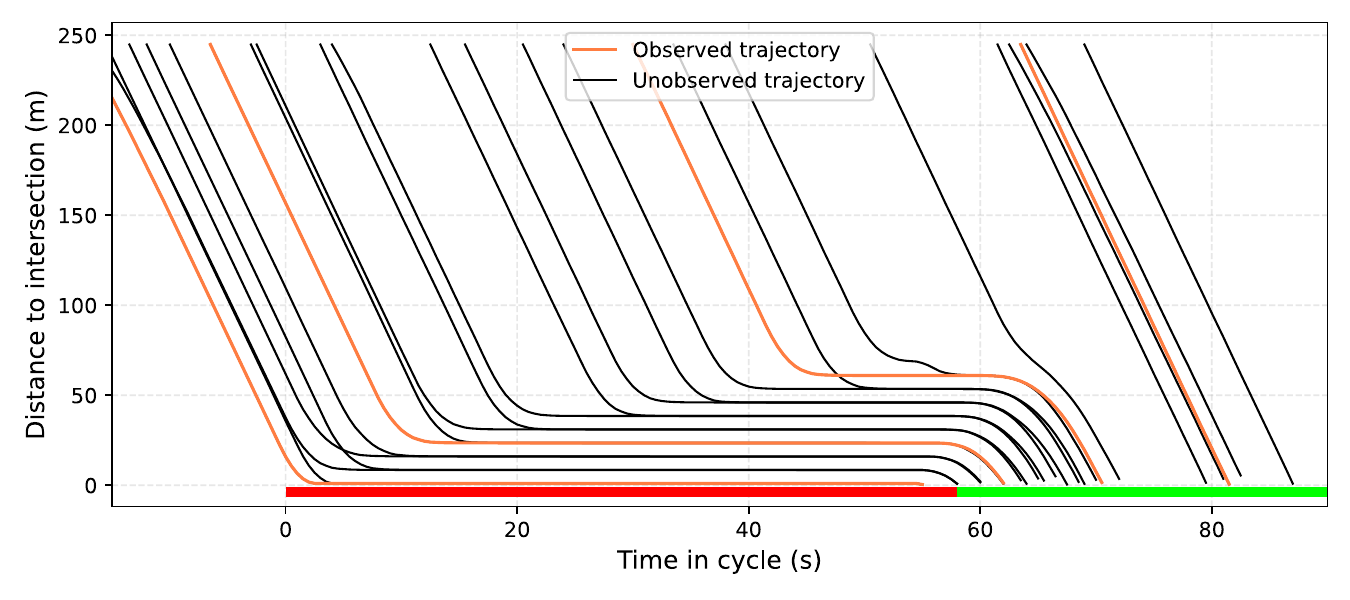}
    \caption{Time-space diagram of a movement. }
    \label{fig:problem-statement}
\end{figure}

This paper focuses on traffic state estimation at signalized intersections with vehicle trajectory data. As shown in Figure \ref{fig:problem-statement}, orange and black lines denote the observable and unobservable vehicle trajectories, respectively. By utilizing the observed trajectory as the input, we propose methods to estimate both real-time traffic state and unknown traffic parameters. The real-time traffic state is described by the queue length at each time while traffic parameters include penetration rate and those related to the vehicle arrival process. In this paper, we also limit the scope to a single movement, which refers to a certain moving direction (e.g., left-turn, through) of a signalized intersection. It is also assumed that this movement uses dedicated lanes and does not share lanes with other movements. 

Similar to many of the existing works as aforementioned, we need to introduce the following Assumption \ref{assump:stationary}, which is the foundation that allows us to aggregate more historical data. 
\begin{assumption}  \label{assump:stationary}
    Within the aggregation period, both the vehicle arrival process and penetration rate are assumed to be stationary. In other words, the penetration rate and those parameters related to the vehicle arrival process stay the same. 
\end{assumption}

The aggregation period in Assumption \ref{assump:stationary} refers to the period within which traffic demand stays relatively stationary in the real world. The common practice to obtain the aggregation period is to split the whole day into different times of day (TODs) such as morning peak hours, mid-day off-peak, and evening peak hours, etc. Each TOD period, which is usually several hours, can be used as an aggregation period. We might further extend aggregation periods to multiple days by assuming that traffic patterns of the same TOD on different days are also similar. However, since the traffic demand is time-varying and changes day by day, Assumption \ref{assump:stationary} becomes less valid if the aggregation period is too long. Consequently, it is a trade-off when determining the duration of the aggregation period: a short aggregation period might not provide sufficient data while a long aggregation period tends to undermine the stationary assumption. 

In this paper, a stochastic binary arrival process is used, and the probability that there is a new vehicle arrival from time $t$ to $t+1$ is given by $a(t)$. To simplify the estimation problem, we use a parametric arrival function $a(t)=a_\mu(t)$ with $\mu$ as the parameter. This paper does not specify the parametric function $a_\mu(t)$, which can be chosen given different scenarios. For example, if the upstream of this movement is not regulated by traffic signals, a uniform arrival is often used, i.e., $a_\mu(t)=\mu$. In this case, the arrival process will converge to a Poisson process if the time interval is close to zero. If the upstream of this movement is controlled by a traffic signal with the fixed cycle $C$ or they are controlled by real-time controllers with an average cycle length of $C$, the following simplification can be made:
\begin{equation}
    a(kC + t) = a_\mu (t), \quad \forall k \in\{0,1,\dots\},\quad \forall t \in\{1, 2,\dots, C\}
\end{equation}
which means that vehicle arrival is cyclic and the arrival profile of each cycle is determined by the parametric function $a_\mu (t), t\in \{1, 2,\dots, C\}$, which can be a polynomial or piece-wise function.

% Note that Assumption \ref{assump:stationary} does not mean that the vehicle arrival does not change with time, it is time-varying but determined by a stationary $\mu$ which stays the same within the aggregation period. 

This paper also assumes that some parameters are known or can be estimated in advance, including saturation flow rate (Supplementary information in \cite{wang2024osaas}), jam density \citep{lloret2023jam}, and even signal timing \citep{axer2017signal,du2019signal}. With this assumption, the main parameters to be estimated in this paper include the vehicle arrival $\alpha_\mu(t)$ and penetration rate $\phi$. Here the penetration $\phi$ refers to the probability that a vehicle can be observed. We also assume that observable vehicles are randomly distributed among all vehicles. 

\section{Model formulation}   \label{sec:formulation}

\subsection{Probabilistic time-space (PTS) model}  \label{sec:pts-model}

The estimation methods are formulated based on the probabilistic time-space (PTS) model proposed by \cite{wang2024osaas}. Here we only provide a brief introduction to this model with similar notations in the original paper. The PTS model assumes that all vehicles follow a homogeneous deterministic Newell's car following model \citep{newell2002simplified}: Each vehicle only has the stop state and the free-flow state with a given free-flow speed $v_f$. This assumption ignores the heterogeneity and randomness of driving behaviors since most of the uncertainty arises from incomplete observation and stochastic traffic demand under a low penetration rate.  

The PTS model also applies a discrete approximation. Let $\Delta t$ be the time interval. The unit traffic flow $\Delta u$ per time interval $\Delta t$ at the saturation flow rate $q^m$ is given by:
\begin{equation}  \label{eq:discrete-unit-flow}
    \Delta u=q^m z\Delta t
\end{equation}
where $z$ denotes the number of lanes of this movement. In this paper, we choose $\Delta t$ such that the unit traffic flow $\Delta u$ equals exactly one vehicle. For example, if we have the saturation flow rate $q^m=1800$ veh$/$(lane $\cdot$ hour) $=$ 0.5 veh$/$(lane$\cdot$sec) and number of lanes $z=2$, $\Delta t$ will be $1$ second such that $\Delta u$ refers to one vehicle. Since the unit traffic flow $\Delta u$ is set as one vehicle in this paper, we will use ``each unit traffic flow $\Delta u$'' and ``one vehicle'' interchangeably. We also use $\Delta t$ as $1$ to simplify the notation. 

\begin{figure}[h!]
    \centering
    \includegraphics[width=0.95\linewidth]{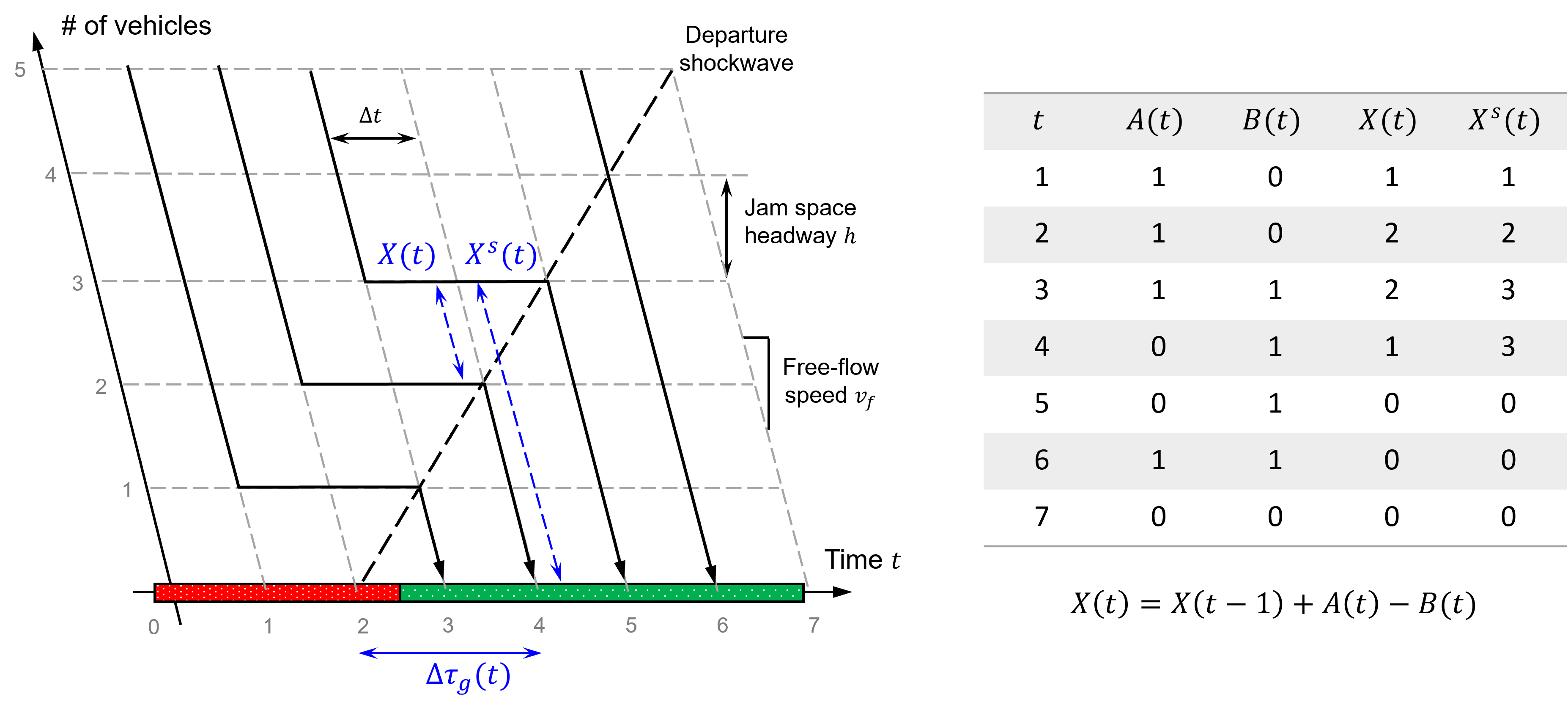}
    \caption{Illustration of the probabilistic time-space (PTS) model based on the Newellian coordinates. Reproduced from \cite{wang2024osaas}}
    \label{fig:pts-model}
\end{figure}

Based on the given assumption and discrete approximation, the PTS model is established based on a new coordinate system called Newellian coordinates \citep{wang2024osaas}, of which the horizontal axis is the free-flow arrival time while the vertical axis is the number of vehicles. As shown in Figure \ref{fig:pts-model}, the spatial-temporal vehicle trajectories can be projected to a point queue process under the Newellian coordinates, and the dynamical equation is given by:
\begin{equation}  \label{eq:queueing-dynamics}
    X(t) = X(t-1) + A(t) - B(t) = X'(t) - B(t)
\end{equation}
where $X(t)$ denotes the number of stopped vehicles right after time $t$, $A(t)$ and $B(t)$ denote the vehicle arrival and departure, respectively. With the discrete approximation, both the vehicle arrival and departure are binary at each time: they are either 0 or one vehicle. 

While $X(t)$ is the number of stopped vehicles at time $t$, $X^s(t)$ denotes the location of the end of the queue. They are also referred to as the point queue and spatial queue, respectively. Under given assumptions stated before, they have a deterministic mapping:
\begin{equation}  \label{eq:spatial-point-queue-mapping}
    X^s(t) = \psi(X(t)) =  X(t) + \Delta \tau_g(t)
\end{equation}
where $\Delta \tau_g$ is the elapsed green time as shown in Figure \ref{fig:pts-model}. We also use $\psi^{-1}$ to represent the inverse of $\psi$, i.e., $X(t) = \psi^{-1}(X^s(t))$.

% \begin{equation}
%     h = \frac{\Delta u \cdot h_0}{z}=q^mh_0\Delta t.
% \end{equation}

By applying a stochastic arrival process, the point queue process given by Equation (\ref{eq:queueing-dynamics}) becomes a stochastic discrete queueing model. Since the arrival $A(t)$ is binary for each time, we assume that it follows a Bernoulli distribution with the arrival probability $\alpha(t)$. That is, $\mathbb{P}(A(t)=1)=\alpha(t)$. The departure $B(t)$ is determined by the following equation:
\begin{equation} \label{eq:departure-prob}
    \mathbb{P}(B(t)=1) = b(t) =\mathbb{P}(X(t) \ge 1 \;\&\; S(t) = 1)
\end{equation}
where $b(t)$ denotes the departure probability while $S(t)$ is the traffic signal states: $1$ and $0$ represent green and red light, respectively. Equation (\ref{eq:departure-prob}) simply means that there will be a departure whenever the queue is not empty ($X(t)\ge 1$) while the traffic signal state is green ($S(t)=1$). 

% The departure process given by Equation (\ref{eq:departure-prob}) is deterministic, which might have certain limitations. In the real world, different factors would contribute to a stochastic departure process, such as a random perception reaction time and the green start-up loss. By ignoring the randomness of the departure process, it will slightly underestimate the system uncertainty.

Let $x(t,k)$ denote the probability that the queue length $X(t)$ is $k$ at time $t$ and $\bm{x}(t,:)$ represent the probability mass function (pmf) of queue length at time $t$. The transition of the queue length distribution is given by: 
\begin{equation}\label{eq:queueing-transition}
\begin{aligned}
    \bm{x}(t+1, :) & = f(\bm{x}(t-1,:), a(t), S(t)) \\ & = \left\{ \begin{aligned}
        x'(t,k+1) &=x(t-1,k)\cdot a(t)+x(t-1,k+1)\cdot (1-a(t)) \\
        x(t,k) &= x' (t,k+1)\cdot S(t)+x' (t,k)\cdot (1-S(t)),\;\forall k\ge 1 \\
        x(t,0) &=x' (t,1)\cdot S(t)+x' (t,0)  \\
        b(t) &=\sum_{k=1}^\infty x'(t,k)\cdot S(t)
    \end{aligned}\right.
\end{aligned}
\end{equation}
where $f(\cdot)$ denotes the transition mapping function. 

The other part of the PTS model is the probabilistic time-space (PTS) diagram, which projects the stochastic point queue process given by Equation (\ref{eq:queueing-dynamics}-\ref{eq:queueing-transition}) back to the spatial-temporal space to obtain the spatial-temporal distribution of vehicle trajectories. Please refer to the Method section of \cite{wang2024osaas} for the detailed derivation. In this way, the PTS model establishes the bidirectional mapping between the point queue process and spatial-temporal vehicle trajectories such that a simple point-queue model can be utilized to describe traffic flow dynamics near intersections. Note that what we have introduced in this subsection does not consider the over-saturated scenario when there is a residual queue at the end of a cycle, the supplementary material of \cite{wang2024osaas} shows how the PTS model can deal with over-saturation. It turns out the over-saturated case (with the residual queue) is similar to the under-saturation case but is more complicated and tedious. Therefore, we will only introduce the simple scenario when there is no residual queue in this paper. This does not mean that the proposed method can only be applied to the simple case. 

\subsection{Trajectory Encoding and Observation Model}  \label{sec:observation-model}

\begin{figure}[h!]
    \centering
       \includegraphics[width=.85\textwidth]{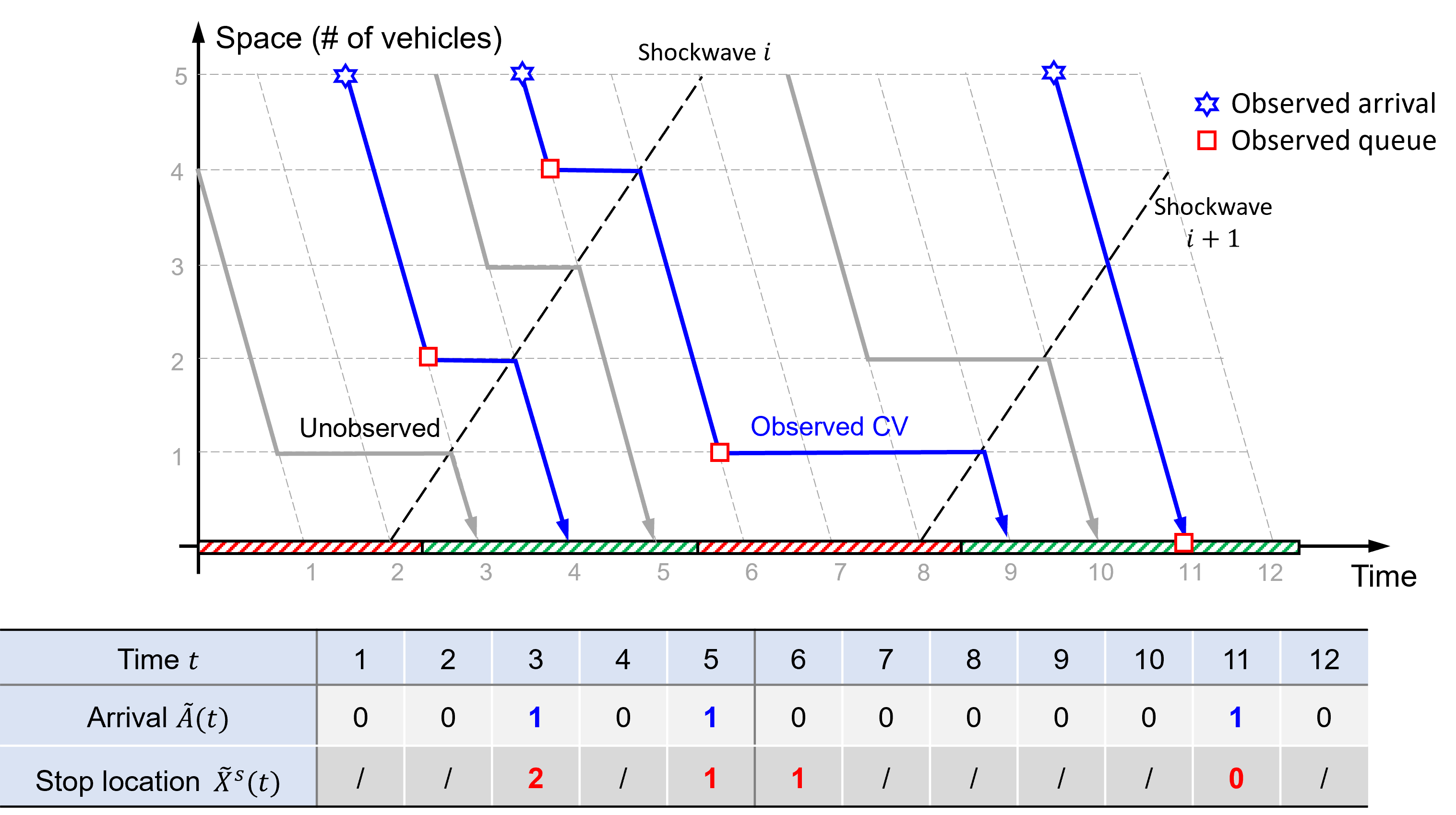} 
    \caption{Observation model and encoding of observed vehicle trajectories. }
    \label{fig:observation-model-encode}
\end{figure}

This subsection introduces the observation model which establishes the connection between the observed vehicle trajectories and the unknown values to be estimated. Figure \ref{fig:observation-model-encode} illustrates the encoding of observed vehicle trajectories. Each observed vehicle trajectory can be encoded as 1) arrival $\tilde{A}(t)$ denoted by the blue star, and 2) stop location $\tilde{X}^s(t)$ denoted by the red rectangle. Notations with the superscript $\sim$ mean that they come from observed vehicle trajectories. 

Here we will show more details about the example given by Figure \ref{fig:observation-model-encode}. As shown in the figure, there are three observed trajectories that arrive at time $t=3$, $t=5$, and $t=11$. Therefore, we have the observed arrival $\tilde{A}(t)=1$ for $t=3,5,11$ and $\tilde{A}(t)=0$ for the rest of the time instants. These three observed trajectories represent different cases:
\begin{enumerate}
    \item The first trajectory arrives at time $3$ is a typical trajectory that stops once before passing the intersection. Other than the observed arrival $\tilde{A}(3)=1$, we also have the observed queue length $\tilde{X}^s(3)=2$ since the trajectory stops at location $2$. 
    \item The second trajectory (arrives at time $5$) is an over-saturated case that stops twice before passing the intersection. For the first stop, we have the observed queue length $\tilde{X}^s(5)=4$. Besides, we also have the second stop $\tilde{X}^s(6)=1$. 
    \item The third trajectory (arrives at time $11$) directly passes the intersection without a stop. Therefore, we have the observed queue length $\tilde{X}^s(11)=0$.
\end{enumerate}

% \begin{remark}
%     Since the PTS model only allows vehicle trajectories to travel on the grid of the Newellian coordinates, real-world trajectories need to be converted to an approximated trajectory that is composed of edges of the Newellian coordinates. This can be easily achieved through different methods. For example, \cite{wang2023trajectory} proposed simple algorithms to split each trajectory into three different states: free flow state, transition state, and stop state. The transition state can be further eliminated by extending the free-flow state and stop state. In this way, the trajectory will only have the free-flow state and stop state. 
% \end{remark}

As we can see from this example, except for the split-failure trajectory that has two stops, each observed arrival $\tilde{A}(t)$ is always associated with one observed stop location $\tilde{X}^s(t)$ at the same time. This is a result of the adoption of the Newellian coordinates, which use the free-flow arrival time as the time. Consequently, the newly arrived vehicles instantaneously join the queue. The observation model can be decomposed into 1) whether a new arrival is observed $p(\Tilde{A}(t)\vert A(t), \phi)$ and 2) what is the observed queue length $p(\Tilde{X}^s(t)\vert X(t), \Tilde{A}(t))$. The observed arrival $p(\Tilde{A}(t)\vert A(t), \phi)$ is determined by:
\begin{subequations} \label{eq:observed-arrivals}
    \begin{align}
       p(\Tilde{A}(t) = 1\vert A(t) = 1, \phi) & = \phi \label{eq:observed-arrival-a}\\
    p(\Tilde{A}(t) = 0\vert A(t) = 1, \phi) & = 1 - \phi \label{eq:observed-arrival-b} \\
    p(\Tilde{A}(t)=0\vert A(t)=0, \phi) & = 1 \label{eq:observed-arrival-c}
    \end{align}
\end{subequations}
where $A(t)=1$ and $A(t)=0$ represent there is and is not an actual new arrival at time $t$ while $\Tilde{A}(t)$ indicates whether we observe this new arrival. $\phi$ is the penetration rate, which is the probability that a vehicle can be observed. Equation (\ref{eq:observed-arrival-a}-{\ref{eq:observed-arrival-b}}) means that we have probability $\phi$ and $1-\phi$ to observe or miss a new arrival, respectively. Equation (\ref{eq:observed-arrival-c}) means that we cannot observe an arrival if there is no new arrival at all. 

\begin{remark}
    To make sure Equation (\ref{eq:observed-arrivals}) holds, each vehicle trajectory needs to exactly correspond to one unit traffic flow $\Delta u$ in the discrete approximation of the PTS model. This can be easily achieved by choosing a proper time interval $\Delta t$ such that $\Delta u$ represents one vehicle (refer to Equation (\ref{eq:discrete-unit-flow}) in Section \ref{sec:pts-model}). Otherwise, one new observed trajectory at time $t$ will not be consistent with $A(t)=1$. For example, if we use a larger $\Delta t$, the unit traffic flow per time step $\Delta u$ will be larger than one vehicle. In this case, if we still regard each observed vehicle trajectory as a complete arrival within the whole time interval $\Delta t$, the arrival rate will be overestimated. 
\end{remark}

For the observed queue length, if we do not observe the new arrival, i.e., $\tilde{A}(t)=0$, then we do not have observed queue length, either:
\begin{equation}  \label{eq:did-not-observe-the-queue}
    p(\tilde{X}^s(t) = \emptyset \mid X(t), \tilde{A}(t) = 0) = 1
\end{equation}
where $\tilde{X}^s(t)=\emptyset$ means that we observe nothing. 

On the other side, whenever we observe a new arrival, i.e., $\tilde{A}(t)=1$, we will also observe the stop location of this vehicle at the same time. Here we assume that the observed stop location is centered by the true spatial queue but with a Gaussian noise term: 
% To simplify the problem, we assume that the observed location is always the true queue length, which is formally stated below:
% \begin{assumption}  \label{assump:accurate-observed-queue}
%     It is assumed that there are no errors or noise in the observed queue length. This means that we assume that 1) there is no noise or errors in the collected GNSS (global navigation satellite system) coordinates; 2) queue lengths of multiple lanes are equivalent; and 3) jam space headway is a deterministic constant. These assumptions can be relaxed by using a noise model; refer to discussions in Section \ref{sec:discussion} for more details. 
% \end{assumption}
% Based on Assumption \ref{assump:accurate-observed-queue}, we have:
% \begin{equation}  \label{eq:observe-the-queue}
%     p(\tilde{X}^s(t) = X^s(t)\mid X(t), \tilde{A}(t) = 1) = 1
% \end{equation}
\begin{equation}\label{eq:observe-the-queue}
    p\left. \left(\tilde{X}^s(t) = X^s(t) + c \right\vert X(t), \tilde{A}(t) = 1\right) = \kappa(c)
\end{equation}
% \begin{subequations}  \label{eq:observe-the-queue}
% \begin{align}
%    & p\left. \left(\tilde{X}^s(t) = X^s(t) + c \right\vert X(t), \tilde{A}(t) = 1\right) = \kappa(c), \quad c=\{- h^g, \dots, 0, 1, \dots, h^g\} \\
%    & p\left. \left(| \tilde{X}^s(t) - X^s(t) | > h^g \right\vert X(t), \tilde{A}(t) = 1\right) = 0
% \end{align}
% \end{subequations}
where $X^s(t) = \psi (X(t))$ is the true spatial queue at time $t$. $\kappa(c)$ is a normalized discrete Gaussian kernel determined as follows:
\begin{subequations}   \label{eq:gauss-kernel}
\begin{align}
    & \kappa(c) = \frac{\mathcal{N}(c; 0, \sigma)}{\sum_{i=-h^g}^{h^g} \mathcal{N}(i;0, \sigma)}  & |c|\le h^g\\
    & \kappa(c) = 0 & |c|>h^g
    \end{align}
\end{subequations}
where $\mathcal{N}(x;0,\sigma)$ is the pmf of the Gaussian distribution with $0$ mean and standard deviation $\sigma$. $h^g$ is the half-width of the Gaussian kernel. 

This kernel function can be used to account for different factors that add noise to the observed queue length, for example, the localization noise of the GNSS (Global Navigation Satellite System). Besides, for the movement with multiple lanes, the queue lengths of different lanes are also different. Since we can only observe the queue length of one lane in most cases, we do not know accurately the queue lengths of the other lanes, although they tend to be similar. Additional assumptions about the lane choice behavior need to be made if we want to establish a more accurate model. Moreover, the stochastic driving behavior, which is ignored when establishing the PTS model, can also be reflected through the randomness of the observed queue length. All these different factors will introduce additional noise to the observed queue length. They all can be incorporated by changing the kernel function in Equation (\ref{eq:gauss-kernel}). Since this will not change the structure of the probabilistic model and the estimation algorithms, we choose not to delve into more details. 

\subsection{Overall Hidden Markov Model (HMM)}

\begin{figure}[h!]
    \centering
       \includegraphics[width=.85\textwidth]{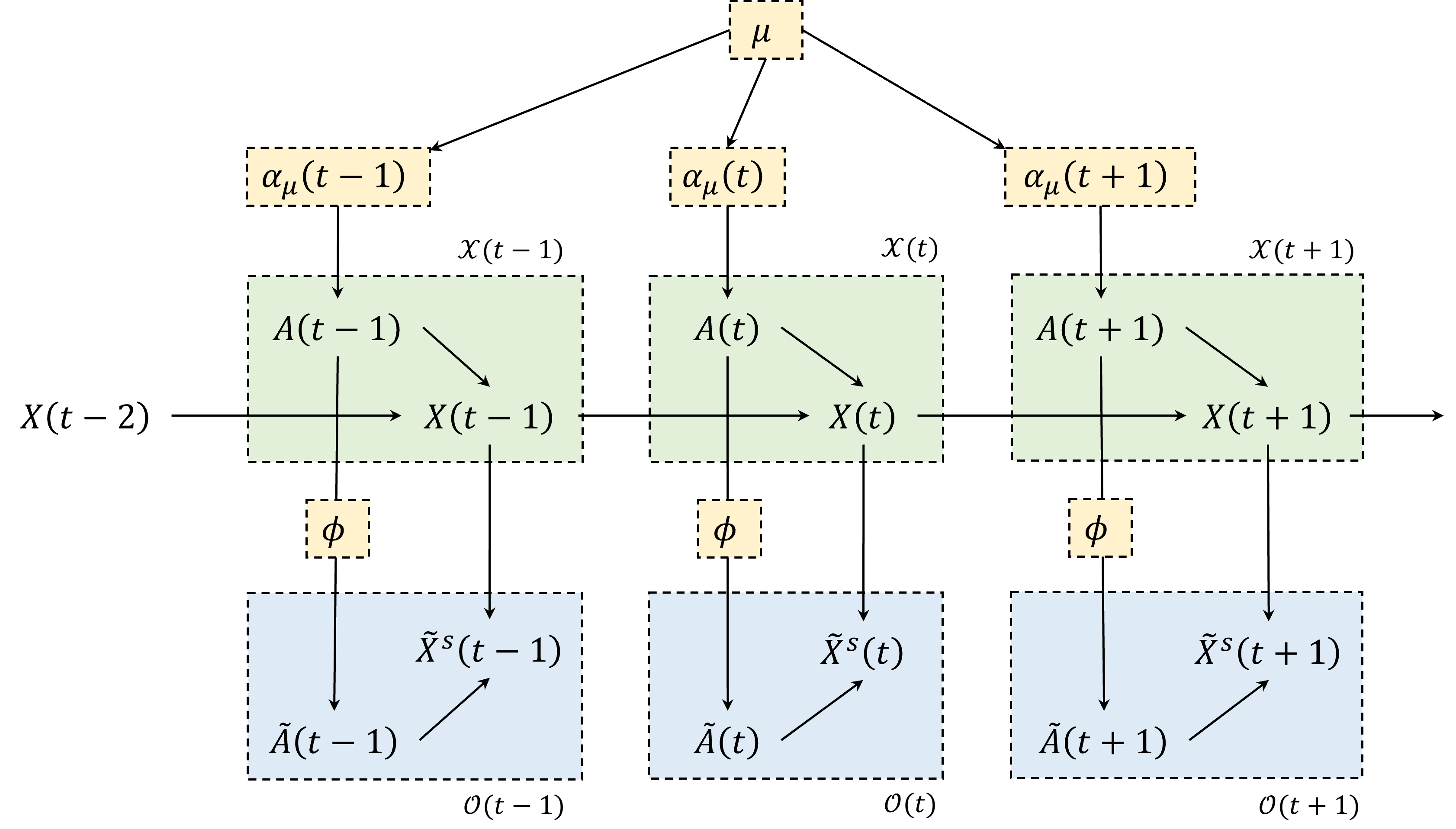} 
    \caption{Overall probabilistic graphical model as a Hidden Markov model. }
    \label{fig:overall-hmm}
\end{figure}

Based on the PTS model (Section \ref{sec:pts-model}) and the observation model (Section \ref{sec:observation-model}), Figure \ref{fig:overall-hmm} shows the probabilistic graphical model (a Bayesian network) of the partially observable system with low penetration rate vehicle trajectory data. The overall Bayesian network is essentially a hidden Markov model (HMM). The hidden state (green color in Figure \ref{fig:overall-hmm}) is the real-time traffic state including the arrival and queue length at each time. Here we use $\mathcal{X}(t)=\{A(t), X(t)\}$ to denote the traffic state at time $t$. The observable state (blue color in Figure \ref{fig:overall-hmm}) consists of the observed arrival and observed queue length at each time, denoted by $\mathcal{Y}(t)=\{\Tilde{A}(t), \tilde{X}^s(t)\}$. Yellow blocks in Figure \ref{fig:overall-hmm} represent the underlying parameters of the hidden Markov model. The arrival probability each time $\alpha_\mu(t)$ is a parametric function with parameter $\mu$, which determines the transition of the hidden traffic state. The penetration rate $\phi$ links the hidden state $\mathcal{X}(t)$ and observation $\mathcal{Y}(t)$. We use $\Theta =\{\mu, \phi \}$ to represent the overall underlying parameters to be estimated. 

Each part of the probabilistic graphical model in Figure \ref{fig:overall-hmm} has been introduced in previous Sections \ref{sec:pts-model}-\ref{sec:observation-model}. Here we will briefly go through the high-level mathematical formulation again so that readers can clearly see how the previous PTS model and observation model construct the overall HMM. The joint distribution of the probabilistic model can be written as:
\begin{equation} \label{eq:general-model}
    p(\mathcal{O}(1:T), \mathcal{X}(1:T), \Theta) = p(\Theta) p(\mathcal{X}(1:T)\vert \Theta)p(\mathcal{O}(1:T)\vert \mathcal{X}(1:T), \Theta)
\end{equation}
where $p(\Theta)$ is the prior of the traffic parameters. In this paper, we use a simple uniform prior, i.e., $p(\Theta)=1$. $p(\mathcal{X}(1:T)\vert \Theta)$ and $p(\mathcal{O}(1:T)\vert \mathcal{X}(1:T), \Theta)$ represent the PTS model and observation model, respectively. The PTS model $p(\mathcal{X}(1:T)\vert \Theta)$ can be further decomposed as:
\begin{equation}  \label{eq:hmm-model-chain}
    p(\mathcal{X}(1:T)\vert \Theta) = p(\mathcal{X}(1)) \prod_{t=1}^Tp(\mathcal{X}(t+1)\vert \mathcal{X}(t), \Theta),
\end{equation}
where $p(\mathcal{X}(1))$ is the initial traffic state at time $1$ and $p(\mathcal{X}(t+1)\vert X(t),\Theta)$ is the transition of the traffic state:
\begin{equation} \label{eq:traffic-model}
    p(\mathcal{X}(t+1)\vert \mathcal{X}(t),\Theta) = p(A(t+1)\vert \alpha_\mu(t+1)) p(X(t+1)\vert X(t), A(t+1), S(t+1))
\end{equation}
where $p(X(t+1)\vert X(t), A(t+1), S(t+1))$ is determined by Equation (\ref{eq:queueing-transition}).

The observation model can be written as:
\begin{equation}   \label{eq:hmm-observation-chain}
     p(\mathcal{O}(1:T)\vert \mathcal{X}(1:T), \Theta) =\prod_{t=1}^T p(\mathcal{O}(t)\vert \mathcal{X}(t), \phi ).
\end{equation}
For each time step, the observation model $p(\mathcal{O}(t)\vert \mathcal{X}(t), \phi )$ can be further decomposed into two parts:
\begin{equation}        \label{eq:observation-equation}
    p(\mathcal{O}(t)\vert \mathcal{X}(t),\phi) = p(\Tilde{A}(t)\vert A(t), \phi) p(\Tilde{X}^s(t)\vert X(t), \Tilde{A}(t)).
\end{equation}
of which the two terms on the right-hand side correspond to Equation (\ref{eq:observed-arrivals}) and Equations (\ref{eq:did-not-observe-the-queue}-\ref{eq:observe-the-queue}), respectively.

% To sum up, this section formulates the probabilistic graphical model as a hidden Markov model (HMM) that connects observation and unknown values to be estimated. In the next section, we will introduce how to estimate both stationary traffic parameters and real-time traffic state based on the HMM. 

\section{Estimation algorithms}   \label{sec:estimation}

\subsection{Estimation problem breakdown}

Based on the HMM formulation in the previous Section \ref{sec:formulation}, this section will introduce the estimation algorithms. The overall estimation problem can be broken into two sub-problems: 1) traffic parameter estimation and 2) traffic state estimation. In this paper, we use the Bayesian approach for both sub-problems such that we have distributional estimation results. 

The Bayesian estimation of traffic parameter $\Theta$ is to find the posterior distribution:
\begin{equation}  \label{eq:parameter-estimation}
    p(\Theta \vert \mathcal{O}(1:T)) = \frac{p(\Theta) p(\mathcal{O}(1:T)\vert \Theta)}{p(\mathcal{O}(1:T))}\propto p(\Theta) p(\mathcal{O}(1:T)\vert \Theta)
\end{equation}
where $p(\Theta)$ is the prior and $p(\mathcal{O}(1:T)\vert \Theta)$ is the likelihood function. However, we do not directly know $p(\mathcal{O}(1:T)\vert \Theta)$ in the HMM formulation. It can be obtained by marginalizing out the hidden state $\mathcal{X}(t)$:
\begin{equation}  \label{eq:marginal-likelihood}
    p(\mathcal{O}(1:T)\vert \Theta) = \int p(\mathcal{O}(1:T),\mathcal{X}(1:T)\vert \Theta) d\mathcal{X}(1:T).
\end{equation}
This is also the reason that $p(\mathcal{O}(1:T)\vert \Theta)$ is often called the marginal likelihood function in HMM. 

Given a certain parameter $\Theta$, traffic state estimation is to find the posterior distribution: $p(\mathcal{X}(1:T)\vert \mathcal{O}(1:T), \Theta)$. In practice, most applications only calculate the following:
\begin{equation} \label{eq:filtering}
    p(\mathcal{X}(t)\vert \mathcal{O}(1:t), \Theta ),\; \forall t.
\end{equation}
In the literature, finding $p(\mathcal{X}(1:T)\vert \mathcal{O}(1:T), \Theta)$ is called smoothing while finding $p(\mathcal{X}(t)\vert \mathcal{O}(1:t), \Theta ), \forall t$ is called filtering. The former one finds the posterior of the hidden state at time $t$ based on all available observations from $1:T$ while the latter one finds the posterior at time $t$ only based on the current and previous observations from $1:t$. The filtering problems only require a forward recursive calculation while the smoothing problems need both forward and backward calculations. In this paper, we also only perform the forward recursive calculation to get the filtering result given by Equation (\ref{eq:filtering}).

Moreover, instead of having a point estimation of the parameter $\Theta$, we have distributional estimation given by Equation (\ref{eq:parameter-estimation}). By incorporating the complete posterior distribution $p(\Theta \mid \mathcal{O}(1:T))$, the real-time traffic state estimation is given below:
\begin{equation}  \label{eq:real-time-filtering}
    p(\mathcal{X}(t)\vert \mathcal{O}(1:t)) = \int \underbrace{p(\mathcal{X}(t)\vert \mathcal{O}(1:t), \Theta )}_{\text{Eq. (\ref{eq:filtering})}} \underbrace{p(\Theta \vert \mathcal{O}(1:T))}_{\text{Eq. (\ref{eq:parameter-estimation})}} d\Theta ,\;\;\forall t.
\end{equation}
Therefore, we need to estimate the parameter first and then substitute the parameter estimation results to get the real-time estimation. 

The rest of this section introduces more details of the estimation algorithms: Section \ref{sec:recursive-bayesian} introduces the recursive algorithm that is used to obtain both the marginal likelihood function $p(\mathcal{O}(1:T)\vert \Theta)$ and real-time estimation $p(\mathcal{X}(t)\vert \mathcal{O}(1:t), \Theta )$ given a certain $\Theta$, at the same time. Based on this recursive algorithm, Section \ref{sec:method-parameter-est} and Section \ref{sec:method-real-time-est} introduce the parameter estimation and real-time traffic state estimation, respectively.

\subsection{Filtering and marginal likelihood calculation} \label{sec:recursive-bayesian}

In this subsection, we will introduce a recursive algorithm that is used to calculate both the marginal likelihood function $p(\mathcal{O}(1:T)\vert \Theta)$ and real-time estimation $p(\mathcal{X}(t)\vert \mathcal{O}(1:t), \Theta )$ at the same time. As a recursive algorithm, we start from $p(\mathcal{X}(t)\vert \mathcal{O}(1:t), \Theta )$ for each time step. The output is the real-time estimation of the next time step: $p(\mathcal{X}(t+1)\vert \mathcal{O}(1:t+1), \Theta)$. Meanwhile, we will keep updating the marginal likelihood function $p(\mathcal{O}(1:t)\vert \Theta)$ for each time $t$. 

By taking $p(\mathcal{X}(t)\vert \mathcal{O}(1:t), \Theta )$ as the input, the real-time traffic state of the next time step $p(\mathcal{X}(t+1)\vert \mathcal{O}(1:t+1), \Theta)$ can be obtained through:
\begin{equation}  \label{eq:update-step}
\begin{aligned}
    p(\mathcal{X}(t+1)\vert \mathcal{O}(1:t+1), \Theta) & =  \frac{p(\mathcal{O}(t+1)\vert \mathcal{X}(t+1), \Theta) p(\mathcal{X}(t+1)\vert \mathcal{O}(1:t), \Theta)}{p(\mathcal{O}(t+1)\vert \mathcal{O}(1:t), \Theta ) } \\
    & \propto p(\mathcal{O}(t+1)\vert \mathcal{X}(t+1), \Theta) p(\mathcal{X}(t+1)\vert \mathcal{O}(1:t), \Theta)
\end{aligned}
\end{equation}
where:
\begin{equation}  \label{eq:prediction-step}
   p(\mathcal{X}(t+1)\vert \mathcal{O}(1:t), \Theta) = \int p(\mathcal{X}(t+1)\vert \mathcal{X}(t), \Theta) p(\mathcal{X}(t)\vert \mathcal{O}(1:t)) d\mathcal{X}(t).
\end{equation}
Equations (\ref{eq:update-step}-\ref{eq:prediction-step}) are widely known as the recursive Bayesian estimation, or Bayes filter. Equation (\ref{eq:prediction-step}) is called the prediction step, which is derived by moving one time step further based on the previous estimation $p(\mathcal{X}(t)\vert \mathcal{O}(1:t))$. Equation (\ref{eq:update-step}) is usually referred to as the update step. Both the Kalman filter \citep{welch1995introduction} and the particle filter \citep{doucet2009tutorial} belong to this recursive Bayesian estimation family. 

In our problem, we need more than the recursive Bayesian estimation given by Equations (\ref{eq:update-step}-\ref{eq:prediction-step}): the parameter $\Theta$ is also unknown and needs to be estimated beforehand. The marginal likelihood function $p(\mathcal{O}(1:T)\vert \Theta)$ can be factorized as follows:
\begin{equation}\label{eq:likelihood-factorization}
    p(\mathcal{O}(1:T)\vert \Theta) =  p(\mathcal{O}(1)\vert \Theta )\prod_{t=1}^{T-1}p(\mathcal{O}(t+1)\vert \mathcal{O}(1:t), \Theta)
\end{equation}
where $p(\mathcal{O}(t+1)\vert \mathcal{O}(1:t)$ can also be obtained through a recursive approach:
\begin{equation}  \label{eq:next-observation}
\begin{aligned}
    p(\mathcal{O}(t+1)\vert \mathcal{O}(1:t), \Theta) = \int & p(\mathcal{X}(t)\vert \mathcal{O}(1:t), \Theta) p(\mathcal{X}(t+1)\vert \mathcal{X}(t), \Theta) \\ & \cdot p(\mathcal{O}(t+1)\vert \mathcal{X}(t+1), \Theta) d\mathcal{X}(t:t+1),
\end{aligned}
\end{equation}
which also takes $p(\mathcal{X}(t)\vert \mathcal{O}(1:t), \Theta)$ as the input. Equation (\ref{eq:next-observation}) is also the denominator of Equation (\ref{eq:update-step}).

In practice, by taking the logarithm of the marginal likelihood function of Equation (\ref{eq:likelihood-factorization}) we calculate the log-likelihood function defined below:
\begin{equation} \label{eq:log-likelihood}
\begin{aligned}
    L(\mathcal{O}(1:T), \Theta) = & \log(p(\mathcal{O}(1:T)\vert \Theta)) \\ = &\log(p(\mathcal{O}(1)\vert \Theta )) + \sum_{t=1}^T \log(p(\mathcal{O}(t+1)\vert \mathcal{O}(1:t), \Theta)).
\end{aligned}
\end{equation}

\begin{algorithm}  
\caption{Recursive calculation of the hidden Markov models} 
\textbf{Input}: Parameter $\Theta$ including arrival rate $\bm{a}_\mu=[a_\mu(1), \dots, a_\mu(T)]$ and penetration rate $\phi$; observed arrival $[\tilde{A}(1), \dots, \tilde{A}(T)]$ and observed queue lengths $[\tilde{X}^n(1), \dots, \tilde{X}^n(T)]$ (Figure \ref{fig:observation-model-encode}); Traffic signal state $\bm{S}=[S(1), \dots, S(T)]$.

\textbf{Initiation:} Initial queue length distribution $\hat{x}(0, k)$ at time $0$ (see Remark \ref{rmk:initial-queue}), initial marginal log-likelihood function $L(0)=0$ where $L(t)=L(\mathcal{O}(1:t)\vert \Theta)$.

\For{$t= 1, 2, ..., T$}{
    Given observed arrival $\Tilde{A}(t)\in\{0, 1\}$, the estimated arrival probability at time $t$:
    \begin{equation}  \label{eq:arrival-posterior}
        \hat{a}(t) = \left\{\begin{array}{cc}
             1  &  \Tilde{A}(t) = 1\\
            \frac{a(t)\cdot (1-\phi)}{1 - a(t)\cdot \phi}  & \Tilde{A}(t) = 0.
          \end{array}\right.
    \end{equation}
    
    \textbf{Prediction step of the queue length distribution}:
    \begin{equation}
        \hat{x}'(t,:) = f(\bm{x}(t-1,:), \hat{a}(t), S(t))
    \end{equation}
    where $f(\cdot)$ is determined by Equation (\ref{eq:queueing-transition}).
    
    \textbf{Update the log-likelihood calculation}:
    \begin{subequations}
        \begin{align}
            % L(t)& =L(t-1) + \log(a(t)\cdot\phi) + \log(\bar{x}(t, \Tilde{X}(t)) &  \hat{a}(t) = 1 \\
            L(t)& =L(t-1) + \log(a(t)\cdot\phi) + \log(K) &  \hat{A}(t) = 1 \\  \label{eq:log-likeli-with-a}
            L(t)& =L(t-1) + \log((1-a(t)) + a(t)\cdot (1-\phi))   &  \hat{A}(t) = 0
        \end{align}
    \end{subequations}
    $K$ in Equation (\ref{eq:log-likeli-with-a}) is calculated as:
    \begin{equation}
        K = \sum_{c=-h^g}^{h^g} \hat{x}'(t,\Tilde{X}(t)+c) \cdot \kappa(c)
    \end{equation}
    where $\Tilde{X}(t)=\psi^{-1}(\Tilde{X}^s(t)))$ is the observed point queue. $\psi(\cdot)$ is the mapping function given by Equation (\ref{eq:spatial-point-queue-mapping}). 
    
    \textbf{Update step of the queue length distribution}:
    \begin{subequations}
        \begin{align}
            % &\hat{x}(t, k) = 1 \;\text{if} \; k=\Tilde{X}(t)\; \text{and}\; 0\; \text{otherwise}\; &\hat{a}(t) = 0 \\
            &\hat{x}(t, k) = \frac{1}{K} \cdot \hat{x}'(t,k) \cdot \kappa(|k - \Tilde{X}(t)|),\quad \forall k  &\hat{A}(t) = 1\\
            &\hat{x}(t, k) = \hat{x}'(t,k),\quad \forall k &\hat{A}(t) = 0 
        \end{align}
    \end{subequations}
}
\textbf{Return}: Marginal likelihood function $L(T)=L(\mathcal{O}(1:T)\vert \Theta)$ and estimated queue length distribution $\hat{x}(t,k),\;\forall t\in\{1,\dots, T\}$ given parameter $\Theta$.
\end{algorithm}

Equations (\ref{eq:update-step}-\ref{eq:log-likelihood}) provide a high-level generic formulation of the overall recursive algorithm for hidden Markov models. The following Algorithm 1 is a detailed implementation of the specific problem in this paper. Given input traffic parameters $\Theta$ and observation $\mathcal{O}(1:T)$, Algorithm 1 outputs 1) the real-time estimation results of the hidden traffic state (filtering): $p(\mathcal{X}(t)\vert \mathcal{O}(1:t), \Theta)$; and 2) the overall marginal log-likelihood given the current input parameters: $L(\mathcal{O}(1:T)\vert \Theta) = \log(p(\mathcal{O}(1:T)\vert \Theta))$. 

\newpage

The derivation of Equation (\ref{eq:arrival-posterior}) in Algorithm 1 is provided as follows: 
\begin{equation}
     p(A(t) = 1 \vert \Tilde{A}(t)=1) = \frac{p(A(t)=1, \Tilde{A}(t)=1)}{p(\Tilde{A}(t)=1)}=\frac{a(t)\cdot\phi}{a(t)\cdot \phi}=1
\end{equation}
which simply means that, whenever we observe a new arrival $\Tilde{A}(t)=1$, we know that the actual vehicle arrival $A(t)=1$. On the other side, when we do not observe any new arrivals, i.e., $\Tilde{A}(t)=0$, we have:
\begin{equation}  \label{eq:predicted-arrival-no-obs}
    \begin{aligned}
    p(A(t) = 1 \vert \Tilde{A}(t)=0) & = \frac{p(A(t) = 1, \Tilde{A}(t)=0)}{p(\tilde{A}(t)=0)} \\
    & = \frac{p(A(t) = 1, \Tilde{A}(t)=0)}{p(A(t) = 1, \Tilde{A}(t)=0) + p(A(t) = 0, \Tilde{A}(t)=0)}\\
     & = \frac{a(t)\cdot (1- \phi)}{a(t)\cdot (1 - \phi) + 1 - a(t) } \\
     & = \frac{a(t)\cdot (1 - \phi)}{1 - a(t)\cdot \phi }
\end{aligned}
\end{equation}

The rest parts of Algorithm 1 are straightforward and follow the main idea given by Equations (\ref{eq:update-step}-\ref{eq:log-likelihood}).

\begin{remark} \label{rmk:initial-queue}
Selection of the initial queue length distribution in Algorithm 1. We can run multiple cycles as a warm-up period to get the initial queue length distribution. A more accurate method is to run multiple warm-up cycles until it reaches a stationary distribution. 
\end{remark}

The following two subsections will introduce the parameter estimation and real-time traffic state estimation, respectively. Both sub-problems are relied on Algorithm 1. 

\subsection{Parameter estimation}  \label{sec:method-parameter-est}

% Based on the marginal log-likelihood function calculated through Algorithm 2, as a frequentist method, the traffic parameter can be estimated through MLE:
% \begin{equation}  \label{eq:traffic-param-mle}
%     \hat{\Theta}_{\text{MLE}} = \argmax_{\Theta} L(\mathcal{O}(1:T), \Theta),
% \end{equation}
% which will provide a point estimation of the traffic parameters. However, this point estimation cannot quantify the uncertainty of the estimated values. Instead of using the frequentist method, the Bayesian method will be used to get the posterior distribution of the unknown parameters $p(\Theta\vert \mathcal{O}(1:T))$ so that we will directly have a distribution of the estimated values.

The Bayesian estimation of the traffic parameters is based on the following Bayes' theorem:
\begin{equation}  \label{eq:bayes-theorem}
    p(\Theta\vert \mathcal{O}(1:T)) = \frac{p(\mathcal{O}(1:T) \vert \Theta) p(\Theta)}{p(\mathcal{O}(1:T) )} \propto p(\mathcal{O}(1:T) \vert \Theta) \cdot p(\Theta)
\end{equation}
where $p(\Theta)$ is the prior of the traffic parameters. We use a prior in this paper, i.e., $p(\Theta)=1$. $p(\mathcal{O}(1:T)\vert \Theta)$ is the marginal likelihood function that can be obtained by Algorithm 1. 

There are different numerical methods to estimate the posterior distribution given by Equation (\ref{eq:bayes-theorem}). For example, if we assume that the arrival process is a homogeneous Poisson process with a single parameter, there are only two parameters (arrival rate $\mu$ and penetration rate $\phi$). In this low-dimensional case, grid search can be used to approximate the posterior distribution: with given ranges and resolutions of both parameters, the parameter space can be split into a mesh grid and the likelihood of each point in the mesh grid can be calculated. Then the posterior of each point can be calculated by normalizing the total probability. 

However, grid search is a time-consuming algorithm. There are many other more advanced and efficient sampling methods such as the importance sampling and Markov Chain Monte Carlo (MCMC) \citep{gelman2013bayesian}. Here we will introduce one specific estimation procedure based on Laplace's approximation and importance sampling. There are four main steps:

\begin{enumerate}
    \item \textbf{Find the MAP} (maximum a posterior). The MAP is given by:
\begin{equation}  \label{eq:traffic-param-map}
    \hat{\Theta}_{\text{MAP}} =\argmax_{\Theta} p(\Theta\vert \mathcal{O}(1:T))
\end{equation}
which is the mode of the posterior distribution. MAP is the same as the maximum likelihood estimation (MLE) if a uniform prior is used.
    \item \textbf{Find the multivariate Gaussian approximation}:
    \begin{equation}
    \hat{\Theta}_{\text{normal}} \sim \mathcal{N} (\hat{\Theta}_{\text{MAP}}, I^{-1}(\mathcal{O}(1:T), \hat{\Theta}_{\text{MAP}}))
\end{equation}
where $I(\cdot)$ is the observed Fisher information matrix:
\begin{equation}
\begin{aligned}
    I(\mathcal{O}(1:T), \hat{\Theta}_{\text{MAP}}) & = - \textbf{Hess}(\log (p(\Theta \vert \mathcal{O}(1:T))))\vert_{\hat{\Theta}_{\text{MAP}}} \\
    & = - \nabla_\Theta \nabla_\Theta^T L(\mathcal{O}(1:T), \Theta) \vert_{\hat{\Theta}_{\text{MAP}}} 
\end{aligned}
\end{equation}
which is the second-order derivative at the MAP. This approximation is also referred to as Laplace's approximation (Chapter 27 in \cite{mackay2003information}. 
    \item \textbf{Apply importance sampling} by utilizing the approximated multivariate Gaussian $\hat{\Theta}_{\text{normal}}$ as the proposal distribution. The multivariate Gaussian $\hat{\Theta}_{\text{normal}}$ does not always provide an accurate approximation, particularly when the data is insufficient. A further step importance sampling can be used to improve the estimation accuracy. We choose not to provide the details of the importance sampling since it is a simple and standard algorithm \citep{gelman2013bayesian}. The output of the importance sampling is a set of sampled points with the corresponding weights: $\{(\Theta_i, w_i), \forall i\}$. 
\end{enumerate}

This procedure is much more efficient than the grid search, which requires much fewer runs of Algorithm 1. However, it might only work well when the multivariate Gaussian provides a good approximation. Otherwise, the following importance sampling in Step 3 will suffer from low efficiency. In this paper, we will use this proposed procedure for the low-dimension scenarios (e.g., uniform Poisson arrival). When the dimensions of the estimation problem go up, for example, a time-varying arrival profile is utilized, we will use the No-U-Turn Sampler (NUTS) \citep{hoffman2014no}, which is a widely used Markov Chain Monte Carlo (MCMC) sampling method. NUTS provides similar outputs to the importance sampling. The main difference is that instead of having a different weight $w_i$ for each sampled point $\Theta_i$, NUTS gives equally weighted samples with $w_i=1$ for all $i$. 
% This can be examined by the approximated effective sample size:
% \begin{equation}
%     S_{\text{eff}} = \frac{1}{\sum_i (\tilde{w}_i)^2}
% \end{equation}
% where $\tilde{w}_i$ is normalized weight given by: $\tilde{w}_i=w_i / \sum_i w_i$.

\subsection{Real-time traffic state estimation}  \label{sec:method-real-time-est}

% Similar to \cite{wang2024osaas}, by assuming that the underlying traffic parameters (i.e., arrival rate and penetration rate) are stationary within a certain time of day (TOD), the overall estimation problem can be decomposed into two parts: 1) stationary parameter estimation and 2) real-time traffic state estimation. 

Based on the results of the importance sampling $\{(\Theta_i, w_i), \forall i\}$, the real-time traffic state estimation is determined by:
\begin{equation}  \label{eq:sample-based-queue-est}
   \hat{\mathcal{X}}^{\text{dis}}(t) \sim p(\mathcal{X}(t)\vert \mathcal{O}(1:t)) = \frac{1}{\sum_i w_i} \sum_{i} w_i\cdot p(\mathcal{X}(t)\vert \mathcal{O}(1:t), \Theta_i )
\end{equation}
which is a weighted sum of all the sampled points according to the associated weights. $\hat{\mathcal{X}}^{\text{dis}}(t)$ denotes real-time estimated traffic state and the superscript ``dis'' means that it is derived by taking the distributional estimated parameter $\Theta$ as input. For each sampled parameter $\Theta_i$, the real-time traffic state estimation $p(\mathcal{X}(t)\vert \mathcal{O}(1:t), \Theta_i )$ is directly available through Algorithm 1. In other words, for each parameter $\Theta_i$, we only need to run Algorithm 1 once, the likelihood function $L(\mathcal{O}(1:T)\vert \Theta_i)$ is used for the previous parameter estimation subsection while the real-time traffic state estimation $p(\mathcal{X}(t)\vert \mathcal{O}(1:t), \Theta_i )$ is used by this subsection. 

However, it is usually not necessary to calculate the real-time traffic state according to Equation (\ref{eq:sample-based-queue-est}), particularly for practical use. By accumulating enough historical data, the point estimator $\hat{\Theta}_{\text{MAP}}$ given by Equation (\ref{eq:traffic-param-map}) would provide a good estimation for parameter $\Theta$ and can be directly used for real-time queue length estimation:
\begin{equation}  \label{eq:queue-est-pt}
    \hat{\mathcal{X}}^{\text{pt}}(t) \sim p(\mathcal{X}(t)\vert \mathcal{O}(1:t), \hat{\Theta}_{\text{MAP}}).
\end{equation}
$\hat{\mathcal{X}}^{\text{pt}}(t)$ also denotes the real-time estimated queue length while taking a point estimation of parameter $\Theta$ as input (with the superscript ``pt'' compared with ``dis'' in Equation (\ref{eq:sample-based-queue-est})). $\hat{\mathcal{X}}^{\text{pt}}(t)$ in Equation (\ref{eq:queue-est-pt}) can be directly obtained by applying $\hat{\Theta}_{\text{MAP}}$ to Algorithm 1. Compared with $\hat{\mathcal{X}}^{\text{dis}}(t)$, $\hat{\mathcal{X}}^{\text{pt}}(t)$ slightly underestimate the uncertainty of the real-time traffic state due to the ignorance of the uncertainty of estimated parameter $\Theta$. However, $\hat{\mathcal{X}}^{\text{pt}}(t)$ is good enough under a low penetration rate since the majority of the uncertainty comes from the sparse observation instead of parameter $\Theta$. 

% Based on Assumption \ref{assump:stationary}, we can choose to accumulate more historical data to improve the estimation accuracy of the parameter $\Theta$. However, this is not helpful for real-time traffic state estimation, which is limited by the sparse observation with a low penetration rate. 

\section{Simulation studies}  \label{sec:simulation}

\subsection{Simulation setup}

We use a simulation environment to validate the proposed methods in this paper. An isolated single movement is built with SUMO \citep{krajzewicz2012recent}. The main parameters of the simulation are available in Table \ref{tab:simulation-steup}. The vehicle arrival is set as a uniform Poisson process with a constant arrival rate. The penetration rate is also a constant, i.e., each vehicle has a pre-determined probability to be observed. Besides, we also run the simulation for 50 cycles as a warm-up. When the vehicle speed is less than $1$ m/s, it is considered as a stopped vehicle. The next two subsections will show the estimation results for traffic parameters and real-time queue length, respectively. 

\begin{table}[h!]
    \centering
    \caption{Simulation setup.}
    \begin{tabular}{c|c}
    \hline
     Parameters    & Values \\
     \hline
     Length & 250 meters \\
     Number of lanes    &  2 \\
     Departure lane    &  Random\\
     Simulation time step    & 0.5 sec\\
     Vehicle arrival    & Uniform Poisson\\
     Cycle length  & 90 sec\\
     Green split & 35 sec\\
     Jam density & 7.5 m/veh \\
     \hline
    \end{tabular}
    \label{tab:simulation-steup}
\end{table}

\subsection{Parameter estimation}

\begin{figure}[h!]
    \centering
    \includegraphics[width=1.0\linewidth]{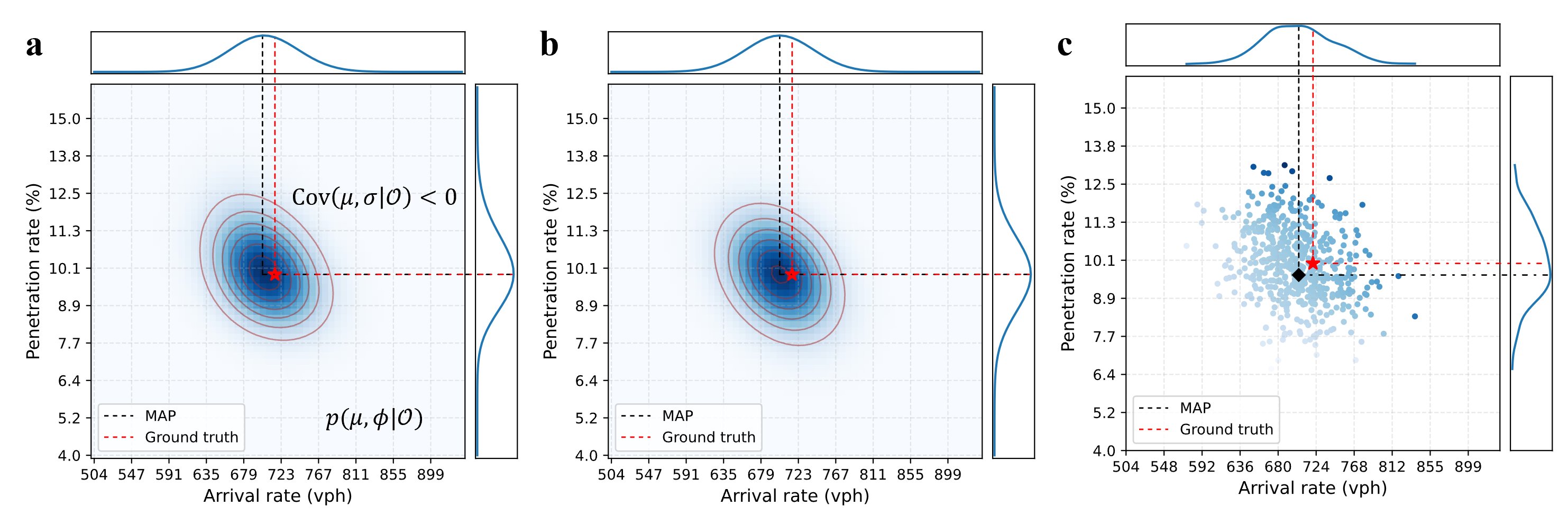}
    \caption{Illustration of the parameter estimation. (a) Posterior obtained through grid search method. (b) Laplace's approximation. (c) Importance sampling based on the Laplace's approximation. }
    \label{fig:estimation-illustration}
\end{figure}

Figure \ref{fig:estimation-illustration} showcases the proposed parameter estimation method, including both the grid search method and the importance sampling method. The same $1$-hour ($40$ cycles) data is utilized for all results in Figure \ref{fig:estimation-illustration}. Figure \ref{fig:estimation-illustration} (a) shows the posterior distribution by using the grid search method. 
% When applying the grid search, the arrival rate ranges from $504$ vph to $936$ vph with an interval of $7.2$ vph while the penetration rate ranges from $4\%$ to $16\%$ with an interval of $0.2\%$. 
A uniform prior is used in this example. The peak value of the distribution (the intersection of two black dashed lines) is the MAP (maximum a posterior), which is used as a point estimation. The red dot is the ground truth. The MAP value is close to the ground truth as shown in the figure. Another observation is that the estimated penetration rate and the arrival rate are negatively correlated, i.e., $\text{cov}(\mu, \phi \vert \mathcal{O})<0$. This is because the production of these two values is approximately the number of observed vehicles, which is a given constant. This means that an overestimation of one of these two parameters might lead to the underestimation of the other one. 

However, as aforementioned in Section \ref{sec:method-parameter-est}, the grid search is time-consuming. The likelihood function needs to be calculated according to Algorithm 1 for each point in the mesh grid to obtain the entire heatmap. Figure \ref{fig:estimation-illustration} (b-c) shows the results of a more efficient approach introduced in Section \ref{sec:method-parameter-est}, which applies the importance sampling by using Laplace's approximation as the proposal distribution. Figure \ref{fig:estimation-illustration} (b) shows the result of Laplace's approximation, which provides a Gaussian distribution to approximate the posterior in Figure \ref{fig:estimation-illustration} (a). Performing Laplace's approximation is much more efficient compared with grid sampling: it only requires getting the peak value (by using optimization-based methods) and estimating the second-order derivative while the grid search method needs to calculate the likelihood of each point in the mesh grid. Given that Laplace's approximation provides a fairly good approximation of the posterior distribution in this case, the importance sampling can be further utilized by using it as the proposal distribution to obtain a more accurate result. Figure \ref{fig:estimation-illustration} (c) shows the result of the importance sampling: each point is a sampled point and the transparency is the associated importance weight. Figures \ref{fig:estimation-illustration} (a-c) also provide the marginal posterior distribution of both parameters aside from their joint distribution. The marginal posterior distribution can be used as the estimation result if we only look into one specific parameter. 

% Laplace's approximation finds the observed Fisher information, which is the second-order derivative at the peak value of the log-likelihood given by Figure \ref{fig:estimation-illustration} (a). A Gaussian distribution is then used to approximate the posterior distribution, of which the center is the peak value and the variance is the inverse of the observed Fisher information. 

% Therefore, Laplace's approximation provides a much cheaper way to get both the estimated value and reasonable metrics (variance) to quantify its uncertainty.

One of the major advantages of the Bayesian approach is that it directly provide distributional estimation results, which explicitly quantify the uncertainty caused by limited available data. Given the distributional estimation results in Figure \ref{fig:estimation-illustration}, the uncertainty of each estimated value (i.e., penetration rate and arrival rate) can be quantified by the credible interval (CI) of the marginal posterior distribution. While there are different methods to calculate the credible interval given the same posterior, here we choose to use the highest density interval (HDI), which refers to the narrowest interval that contains the required percentage (e.g., 95\%) of the posterior distribution. 

\begin{figure}[h!]
    \centering
    \includegraphics[width=0.95\linewidth]{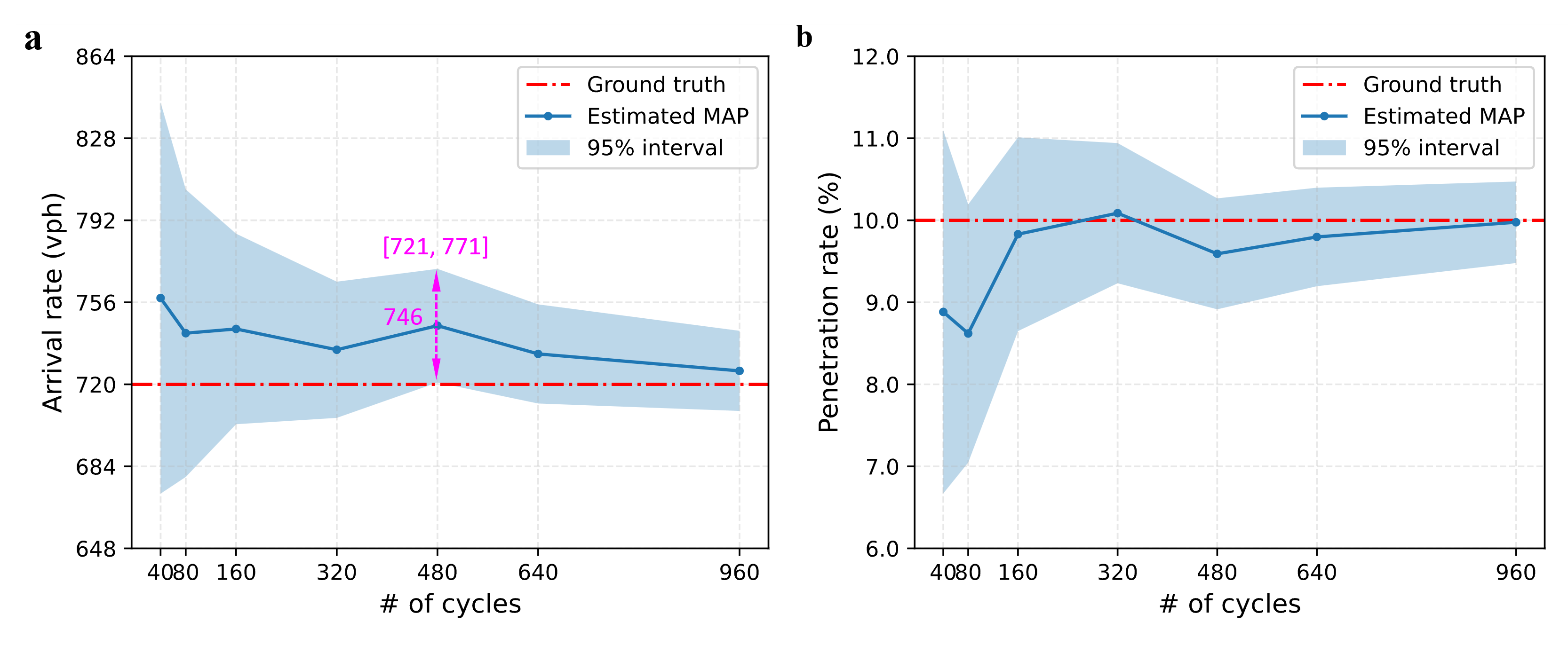}
    \caption{Parameter estimation with uncertainty quantification}
    \label{fig:parameter-marginal}
\end{figure}

Figure \ref{fig:parameter-marginal} shows how the estimated penetration rate and arrival rate change with different amounts of input data, in a specific simulation test. The horizontal axis is the amount of data that is utilized while the vertical axis is the value of the corresponding parameter. The red dashed lines denote the ground truth. The blue lines and regions are estimated results: the blue lines denote the MAP while the blue regions denote the $95\%$ credible intervals. As expected, the credible interval becomes narrower with the increase of available data. Specifically, as illustrated by the pink annotation in Figure \ref{fig:parameter-marginal} (a), when $12$-hour data ($480$ cycles) is utilized, the credible interval of the arrival rate is $[721, 771]$. This can be roughly interpreted as such: by utilizing $12$-hour data, we have a $95\%$ belief that the arrival rate is between $721$ vph and $771$ vph, which is approximately $\pm7\%$ of the MAP ($746$ vph).

To validate the observed credible interval, we run the simulation tests multiple times and check the percentage that the ground truth lies in the credible interval. For the specific case in Figure \ref{fig:parameter-marginal}, the credible interval denoted by the blue area well covers the ground truth denoted by the red dashed line for most parts except for the arrival rate with 480 cycles' data. In this paper, we define coverage rate (CR) as the percentage of the tests that the ground truth is within the obtained credible interval. If the observed data exactly follows the probabilistic model that is used to obtain the likelihood function, the CR will converge to the pre-determined percentage of the credible interval as the number of tests increases. The average width of the credible interval (AWCI) given a certain percentage (e.g., $95\%$) is used to quantify the uncertainty of the estimated values. Intuitively, a larger AWCI indicates a flatter posterior distribution as well as a larger uncertainty. The mean absolute percentage error (MAPE) is utilized to evaluate the accuracy of the point estimation (MAP). Given the point estimation $\hat{\theta}$ of an unknown parameter with true value $\theta$, the MAPE is defined as:
\begin{equation}
    \text{MAPE} = \frac{1}{N}\sum_{i=1}^N \left\lvert \frac{\hat{\theta} - \theta}{\theta}\right\rvert \times 100 \%. 
\end{equation}

Table \ref{tab:result-tab-1} shows the results of the parameter estimation (including both the arrival rate and penetration rate) given different amounts of data ranging from 1 hour to 16 hours. Here we use the duration of the aggregation period to denote the amount of data that is utilized. For example, ``12 hours'' means that the vehicle trajectory utilized for estimation is collected from a total of 12 hours in real-world time. As aforementioned in this paper, the aggregation time can come from different days as long as we believe that the traffic demand is relatively stationary. 12 hours could account for 6 days' data if the time of day period is 2 hours each day (e.g., 8:00 -- 10:00 AM). In this paper, we just set the pre-determined parameters to be exactly stationary. Each row in Table \ref{tab:result-tab-1} reports the results of a given duration of the input data, and the results include the MAPE of both estimated values, as well as AWCI and CR under different CI percentages. 500 experiments are repeated to generate the results for each row. For this specific test scenario, the arrival rate (i.e., traffic volume) is $720$ vehicle per hour (vph) and the penetration rate is $10\%$. 

\begin{table}[h!]
\caption{Bayesian parameter estimation results under different sizes of input data and different percentages of credible intervals (test scenario: arrival rate $\mu = 720 $ vph, penetration rate $\phi = 10\%$).  }
\centering
\begin{tabular}{cc|c|cc|cc|cc}
\hline
\multirow{2}{*}{} &  & \multirow{2}{*}{MAPE (\%)} & \multicolumn{2}{c|}{75\% CI} & \multicolumn{2}{c|}{85\% CI} & \multicolumn{2}{c}{95\% CI} \\
                          &             &     & AWCI  & CR (\%)   & AWCI  & CR (\%)  & AWCI  & CR (\%)  \\ \hline
\multirow{2}{*}{1 hour (40 cycles)}   & $\mu$ (vph)  & 5.5 & 104 & 70.4 & 128 & 80.6 & 171 & 86.6 \\
                          & $\phi$ (\%) & 9.8 & 3.0 & 79.0 & 3.7 & 82.6 & 5.0 & 89.6 \\ \hline
\multirow{2}{*}{2 hours (80 cycles)}  & $\mu$ (vph) & 4.0 & 74  & 72.6 & 91  & 80.2 & 122 & 90.8 \\
                          & $\phi$ (\%) & 7.0 & 2.1 & 74.8 & 2.6 & 83.8 & 3.5 & 95.4 \\ \hline
\multirow{2}{*}{4 hours  (160 cycles)}  & $\mu$ (vph) & 2.7 & 53  & 73.2 & 65  & 79.6 & 87  & 91.8 \\
                          & $\phi$ (\%) & 4.9 & 1.5 & 75.0 & 1.8 & 83.8 & 2.5 & 94.2 \\ \hline
\multirow{2}{*}{8 hours  (320 cycles)}  & $\mu$ (vph) & 1.9 & 37  & 73.8 & 46  & 82.0 & 61  & 93.0 \\
                          & $\phi$ (\%) & 3.6 & 1.0 & 71.2 & 1.3 & 83.6 & 1.7 & 93.6 \\ \hline
\multirow{2}{*}{12 hours  (480 cycles)} & $\mu$ (vph) & 1.6 & 31  & 73.4 & 37  & 81.2 & 50  & 92.2 \\
                          & $\phi$ (\%) & 2.9 & 0.9 & 69.0 & 1.0 & 85.2 & 1.4 & 94.6 \\ \hline
\multirow{2}{*}{16 hours  (640 cycles)} & $\mu$ (vph) & 1.4 & 26  & 73.2 & 32  & 82.8 & 44  & 94.4 \\
                          & $\phi$ (\%) & 2.6 & 0.7 & 70.2 & 0.9 & 80.8 & 1.2 & 92.6 \\ \hline
\end{tabular}
\label{tab:result-tab-1}
\end{table}

% \begin{table}[]
% \caption{Bayesian parameter estimation results under different sizes of input data and different percentages of credible intervals (test scenario: arrival rate $\mu = 720 $ vph, penetration rate $\phi = 10\%$, 500 experiments).  }
% \centering
% \begin{tabular}{cccccccc}
% \hline
% \multirow{2}{*}{Est. on $\mu$ (vph)} & \multirow{2}{*}{MAPE (\%)} & \multicolumn{2}{c}{75\% CI} & \multicolumn{2}{c}{85\% CI} & \multicolumn{2}{c}{95\% CI} \\
%                        &     & AWCI & CR (\%) & AWCI & CR (\%) & AWCI & CR (\%) \\ \hline
% 1 hour (40 cycles)     & 5.5 & 104  & 70.4    & 128  & 80.6    & 171  & 86.6    \\
% 2 hours (80 cycles)    & 4.0 & 74   & 72.6    & 91   & 80.2    & 122  & 90.8    \\
% 4 hours  (160 cycles)  & 2.7 & 53   & 73.2    & 65   & 79.6    & 87   & 91.8    \\
% 8 hours  (320 cycles)  & 1.9 & 37   & 73.8    & 46   & 82.0    & 61   & 93.0    \\
% 12 hours  (480 cycles) & 1.6 & 31   & 73.4    & 37   & 81.2    & 50   & 92.2    \\
% 16 hours  (640 cycles) & 1.4 & 26   & 73.2    & 32   & 82.8    & 44   & 94.4    \\ \hline
% \end{tabular}
% \end{table}

As shown in Table \ref{tab:result-tab-1}, both the MAPE and AWCI decrease with the increase of the input data. The CRs are close to the pre-determined percentages of the credible intervals for all three different values from 75\% to 95\%, which demonstrates the effectiveness of the observed credible intervals. However, readers might notice that CRs are slightly smaller than the given CI percentages. There are several possible reasons. First, the estimator might be slightly biased since it is hard to set all pre-determined parameters (e.g., free flow speed, average start-up loss time, and saturation flow rate, etc.) to be perfectly accurate. If the estimator is biased, the CR will be less than the given CI percentage. Second, we apply different simplifications and assumptions when we establish the probabilistic model, these different approximations might cause slight overestimation or underestimation of the system uncertainties. Even though, the CRs in Table \ref{tab:result-tab-1} are still close to the CI percentages. In summary, Table \ref{tab:result-tab-1} indicates that the proposed method can not only provide a good point estimation of unknown traffic parameters but also effectively quantify the estimation uncertainties.

% One of the major advantages of the Bayesian estimation method is that it can not only provide the estimated value but also the distribution as well as the associated uncertainty. Figure \ref{fig:parameter-uncertainty} shows the estimation uncertainty of the traffic parameters under different conditions. The RSD is used to quantify the uncertainty of the estimated value, which is determined by the standard derivation divided by the mean value (unit: \%). For each of the figures, the horizontal axis is the number of input cycles while the vertical axis is the RSD of the estimated parameters. Generally, the estimation uncertainty decreases with the increase of the input data. The left figures show how different parameters influence the arrival rate estimation while the right figures are about the penetration rate estimation. 

Table \ref{tab:param-est-2} shows more results under different penetration rates and arrival rates. The duration of the input data is fixed at 8 hours. The simulation experiment is also repeated 500 times for each parameter setting. Intuitively, both MAPE and AWCI decrease with the increase in the penetration rate, which means that the estimation becomes more accurate with less uncertainty when the penetration goes higher. On the other side, both MAPE and AWCI decrease when the traffic volumes become larger. This is because we will have more vehicle delays and stops with a larger traffic volume (under the same signal timing parameter), and stopped vehicles contain more useful information for the estimation of traffic parameters.

\begin{table}[h!]
\caption{Bayesian parameter estimation results under different arrival rates and penetration rates. Each experiment uses 8-hour data as input. (CR: \%, MAPE: \%, AWCI has the same unit with the estimated values.)  }
\centering
\begin{tabular}{cl|ccc|ccc|ccc|ccc}
\hline
\multicolumn{1}{l}{\multirow{2}{*}{}} &
   &
  \multicolumn{3}{c|}{$\mu = 180 $} &
  \multicolumn{3}{c|}{$\mu = 360 $} &
  \multicolumn{3}{c|}{$\mu = 540 $} &
  \multicolumn{3}{c}{$\mu = 720 $} \\
\multicolumn{1}{l}{}         &        & AWCI  & CR   & MAPE & AWCI  & CR   & MAPE & AWCI  & CR   & MAPE & AWCI  & CR   & MAPE \\ \hline
\multirow{2}{*}{$\phi=5\%$}  & $\mu$  & 111 & 94.4 & 14.0 & 96  & 96.0 & 5.7  & 85  & 92.4 & 3.6  & 78  & 94   & 2.3  \\
                             & $\phi$ & 3.4 & 94.8 & 14.7 & 2.1 & 95.4 & 8.4  & 1.5 & 95.4 & 6.3  & 1.3 & 92.8 & 5.1  \\ \hline
\multirow{2}{*}{$\phi=10\%$} & $\mu$  & 78  & 97.6 & 9.9  & 71  & 94.4 & 3.9  & 64  & 92.3 & 2.8  & 61  & 93.0 & 1.9  \\
                             & $\phi$ & 4.3 & 95.8 & 10.6 & 2.8 & 92.2 & 5.9  & 2.1 & 92.6 & 4.3  & 1.7 & 93.6 & 3.6  \\ \hline
\multirow{2}{*}{$\phi=15\%$} & $\mu$  & 61  & 98.0 & 7.7  & 59  & 95.6 & 3.3  & 55  & 94.7 & 2.3  & 53  & 95.0 & 1.6  \\
                             & $\phi$ & 4.8 & 97.8 & 7.5  & 3.3 & 94.2 & 4.8  & 2.5 & 91.9 & 3.5  & 2.1 & 94.2 & 2.7  \\ \hline
\multirow{2}{*}{$\phi=25\%$} & $\mu$  & 45  & 98.8 & 5.7  & 48  & 95.0 & 2.8  & 47  & 91.0 & 2.0  & 46  & 93.0 & 1.3  \\
                             & $\phi$ & 5.1 & 98.0 & 5.2  & 3.8 & 94.6 & 3.4  & 2.9 & 87.0 & 2.1  & 2.5 & 93.5 & 2.2  \\ \hline
\multirow{2}{*}{$\phi=50\%$} & $\mu$  & 32  & 98.2 & 3.2  & 37  & 95.8 & 2.3  & 40  & 97.1 & 1.5  & 42  & 71.6 & 1.9  \\
                             & $\phi$ & 5.3 & 96.2 & 2.8  & 4.3 & 95.0 & 2.3  & 3.5 & 93.2 & 1.5  & 2.9 & 82.0 & 1.8  \\ \hline
\end{tabular}
\label{tab:param-est-2}
\end{table}

\subsection{Real-time queue length estimation}

It is much easier to perform the real-time queue length estimation given the estimated traffic parameters. As introduced in Section \ref{sec:estimation}, a standard recursive Bayesian estimation algorithm (filtering) can be applied to get the posterior distribution of the real-time queue length. Since the real-time estimation part is less challenging, we choose not to delve into details. We mainly show how it works and its general performance. The true values of the traffic parameters (penetration rate and arrival rate) are directly used in this subsection.  

\begin{figure} [h!]
    \centering
    \includegraphics[width=0.65\linewidth]{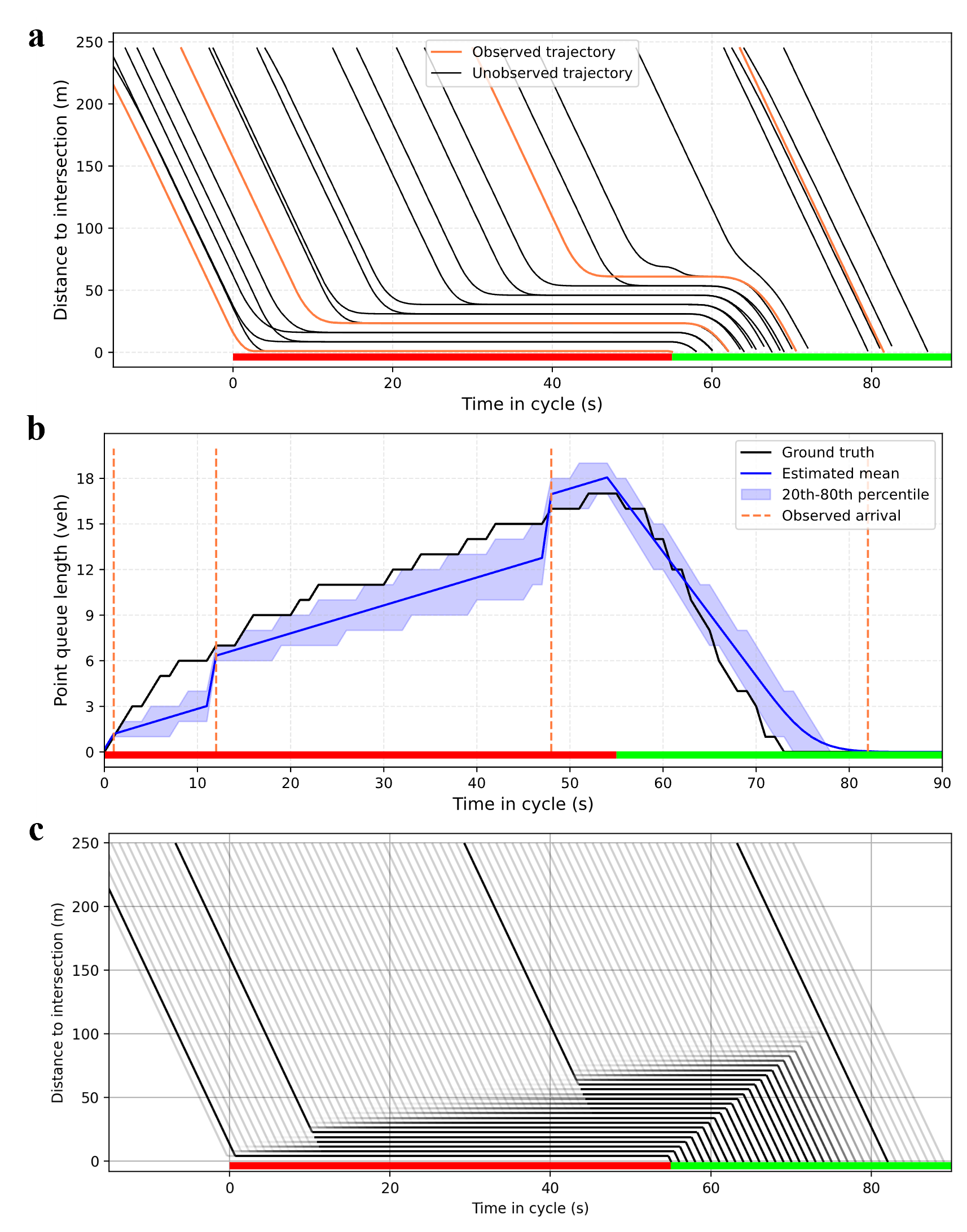}
    \caption{Illustration of real-time queue length estimation.}
    \label{fig:realtime-example}
\end{figure}

Figure \ref{fig:realtime-example} is an illustration of the real-time queue length estimation. Figure \ref{fig:realtime-example} (a) shows the time-space diagram including both observed and unobserved trajectories, denoted by orange and black lines, respectively. Figure \ref{fig:realtime-example} (b) shows the real-time estimated queue length while Figure \ref{fig:realtime-example} (c) is the resulting PTS diagram. Note that the queue length in Figure \ref{fig:realtime-example} (b) is not the spatial queue in common sense (end of the last stopped vehicle) but the converted point queue in units of number of vehicles. Please refer to Section \ref{sec:pts-model} for more details of the relation between the point and spatial queues. Figure \ref{fig:realtime-example} (b) provides a good illustration of how the recursive Bayesian estimation works. The vertical yellow dashed line denotes the time instant in which there is an observed arrival as well as the observed queue length. Any time steps except for these yellow dashed lines mean that we do not have any observation. When there is no observation, the predicted queue length simply evolves according to the given prior arrival rate. In this case, the predicted arrival rate is given by Equation (\ref{eq:predicted-arrival-no-obs}), which is less than the prior arrival rate. When there is an observed new arrival denoted by the orange dashed line, the queue length is updated according to the observed stop location. There is a discontinuous ``jump'' in the estimated queue length profile since the model-predicted queue length has a much larger variance than the newly observed queue. Consequently, the resulting estimated queue length will directly jump to the value around the observed queue length. This discontinuity only happens when we use a filtering algorithm that only performs a forward calculation and updates the current queue length given all previous observations. We will get a more smoothing queue profile if the smoothing algorithm is utilized. Figure \ref{fig:realtime-example} (c) shows the corresponding probabilistic time-space (PTS) diagram, which is derived based on the point queue distribution in Figure \ref{fig:realtime-example} (b) and displays the spatial-temporal distribution of vehicle trajectories.

% As shown in Figure \ref{fig:realtime-example} (b), whenever a new vehicle is observed, the queue length is updated. There could be a discontinuous ``jump'' of the estimated queue length when a new observation is used to update it. This discontinuity is caused by the filtering algorithm, which only finds the posterior of the queue length given all previous observations, i.e., $p(X(t)\vert \mathcal{O}(1:t))$. This can be improved if a smoothing algorithm is applied instead. 

Although the complete queue profile can be obtained through the recursive Bayesian estimation program, we only investigate the maximum queue length for the performance evaluation. The maximum point queue length mostly happens at the start of the green time. Therefore, we directly compare the estimated queue length with the ground truth at the start of the green time for each cycle. We also use the MAPE as the evaluation metric. Let $X^\text{max}_i$ and $\hat{X}^\text{max}_i$ denote the ground truth and estimated maximum queue of cycle $i$. The MAPE is calculated according to:
\begin{equation}  \label{eq:maximum-queue-mape}
    \text{MAPE} = \frac{1}{N} \sum_{i=1}^N \left\vert \frac{X^\text{max}_i - \hat{X}^\text{max}_i}{X^\text{max}_i} \right\vert.
\end{equation}

\begin{figure}[h!]
    \centering
    \includegraphics[width=0.55\linewidth]{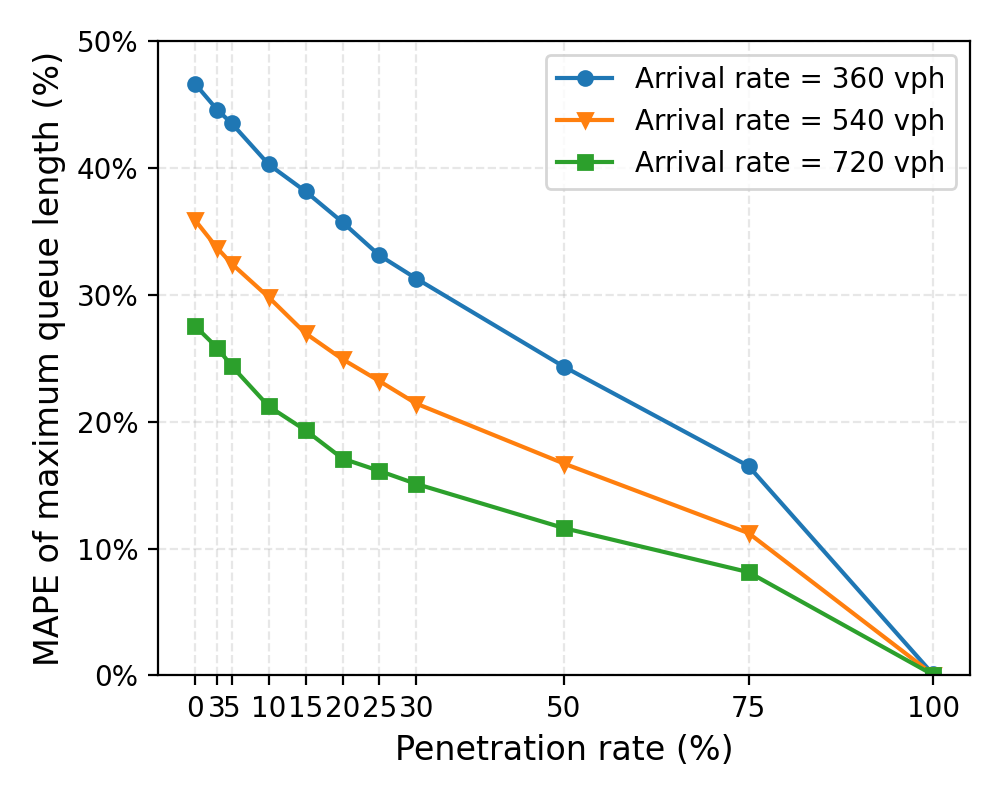}
    \caption{MAPE of maximum queue length estimation under different penetration rates and
arrival rates}
    \label{fig:real-time-results}
\end{figure}

Figure \ref{fig:real-time-results} shows the MAPE of the maximum queue length estimation under different arrival rates and penetration rates. Each point in Figure \ref{fig:real-time-results} uses 75-hour (real-world time, 3000 cycles in total) data generated by the simulation environment. The MAPE monotonically decreases with the increase in both the penetration rate and arrival rate. When the penetration rate is 0, which means there is no observation, the real-time estimated queue length will simply be the average queue length derived based on the given prior arrival rate. Intuitively, a higher penetration rate is always beneficial.  Moreover, when the traffic volume becomes larger, more vehicle trajectories can be observed, which can also improve the accuracy of the queue length estimation. Note that many of the assumptions and simplifications utilized in this paper hold better under a low penetration rate but become less effective when the penetration rate gets higher. This might be one of the reasons we observe fewer improvements in Figure \ref{fig:real-time-results} when the penetration rate goes higher. When the penetration rate is 100\%, we have a complete view of all trajectories, and the estimation error is eliminated.

% However, as shown in Figure \ref{fig:real-time-results}, the MAPE does not decrease to $0$ when the penetration rate is $100\%$. Ideally, the estimation error should be completely eliminated with a $100\%$ penetration rate since all vehicles are observable in this case. This demonstrates the limitation of the proposed estimation method: it is more suitable for low-penetration scenarios.  More discussion is available in Section \ref{sec:discussion}, which also provides possible solutions to improve some of the assumptions.

Unlike traffic parameter estimation which can benefit from aggregating more historical data, real-time estimation highly relies on how much streaming data we can get. Historical data can only provide good prior information for real-time traffic state estimation. If the intersection is under-saturated, the observed vehicle trajectory from the previous cycle becomes immediately useless for the traffic state estimation of the current cycle since the traffic states across different cycles are independent if there is no residual queue. 

\section{Case study with real-world data}  \label{sec:real-world}

\begin{figure}[h!]
    \centering
    \includegraphics[width=0.95\linewidth]{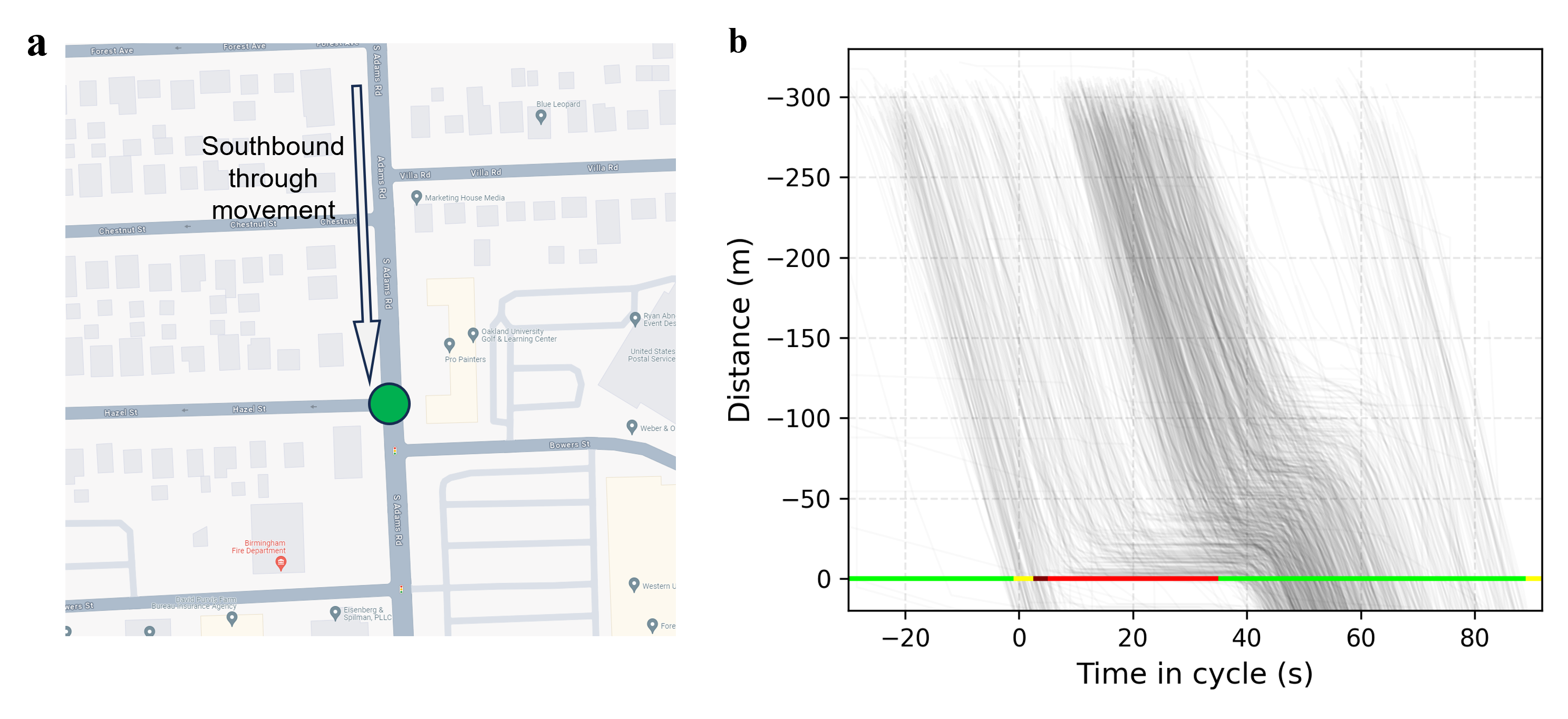}
    \caption{Real-world case study: (a) Intersection geometry. (b) Aggregated time-space diagram. }
    \label{fig:case-study}
\end{figure}

This section provides a case study with real-world data. The vehicle trajectory data is collected from General Motors (GM) vehicles \citep{wang2024osaas}. Figure \ref{fig:case-study} (a) provides the geometry of the studied intersection: Hazel St./Adams Rd., which is located in the City of Birmingham, Michigan. We only investigate the southbound through movement of this intersection as labeled in the figure. The data was collected from 03/07/2022 to 03/11/2022, which were five continuous weekdays. Here we focus on the mid-day period (10:00 -- 15:00) and assume that the underlying traffic parameters remain stationary. Figure \ref{fig:case-study} (b) shows the so-called aggregated time-space diagram, which is obtained by plotting all available data to the same cycle. 

\begin{figure}[h!]
    \centering
    \includegraphics[width=0.95\linewidth]{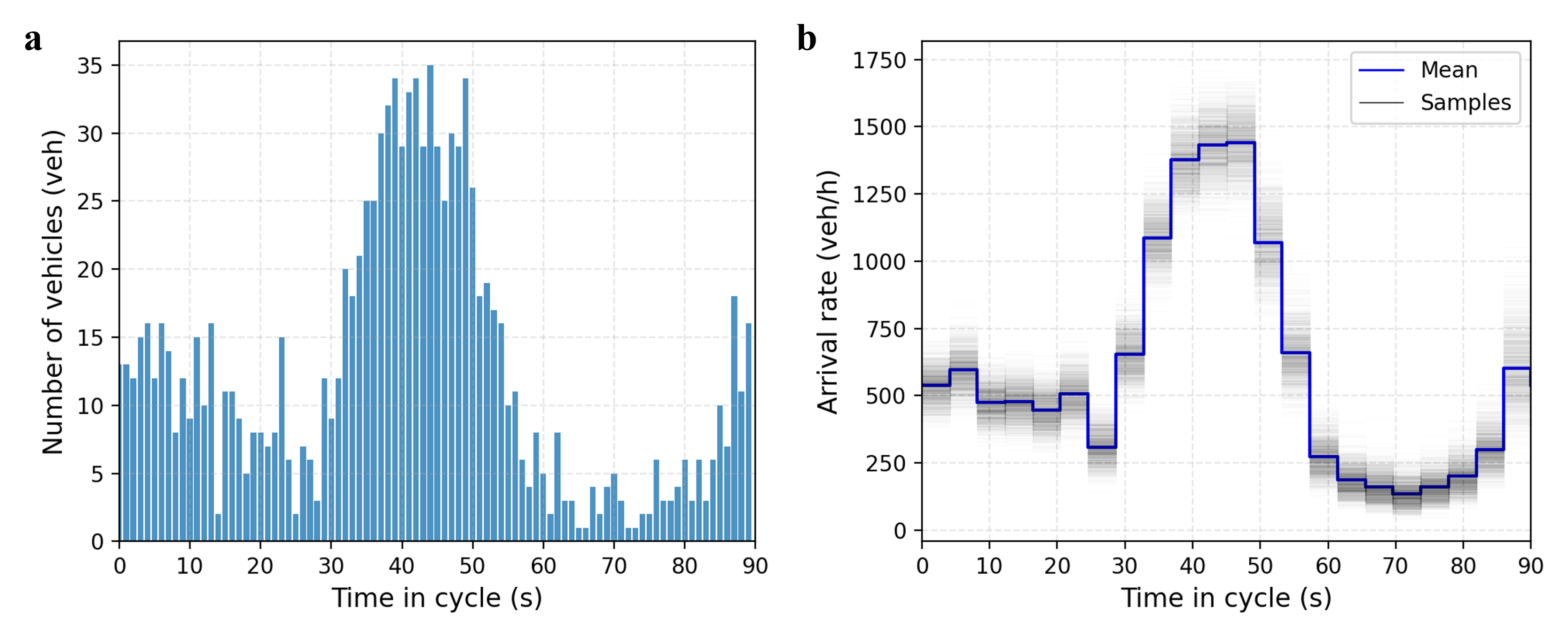}
    \caption{Observed arrival histogram and estimated arrival profile with uncertainties. }
    \label{fig:case-arrival-est}
\end{figure}

Figure \ref{fig:case-arrival-est} (a) shows the arrival time histogram within the cycle that is directly extracted from the aggregated time-space diagram in Figure \ref{fig:case-study} (b). For this specific movement, the upstream arrival is no longer uniform: it is regulated by the upstream intersection. Therefore, we use a piece-wise constant arrival profile, that is:
\begin{equation}
    a_\mu(t) = \mu_i,\; t\in [t_i, t_{i+1}) \quad\forall i. 
\end{equation}

In this problem, the number of parameters is the number of intervals in the arrival profile plus one (penetration rate), which is much larger than the simple uniform arrival case. To ensure a good sampling efficiency, we utilize the No-U-Turn Sampler (NUTS) \citep{hoffman2014no} implemented by PyMC \citep{abril2023pymc}. To set up the NUTS, we use 18 independent Markov chains and each of them has a total of 1000 samples excluding the warm-up period at the beginning. Figure \ref{fig:case-arrival-est} (b) shows the sampling results of the arrival profiles. The black lines show all sampled arrival profiles while the blue line is the expected value. Generally, the blue line in Figure \ref{fig:case-arrival-est} (b) fits the observation in Figure \ref{fig:case-arrival-est} (a) very well. Figure \ref{fig:case-arrival-est} (b) also explicitly shows the uncertainty of the estimated arrival rate of each interval. Intuitively, we will get a denser posterior for the arrival rate of an interval if the arrival remains more steady and more data is available. 

% As shown in Figure \ref{fig:case-arrival-est} (b) samples in Interval A are more concentrated while samples in Interval B are more dispersed, which is consistent with the intuition just stated. 

\begin{figure}[h!]
    \centering
    \includegraphics[width=1\linewidth]{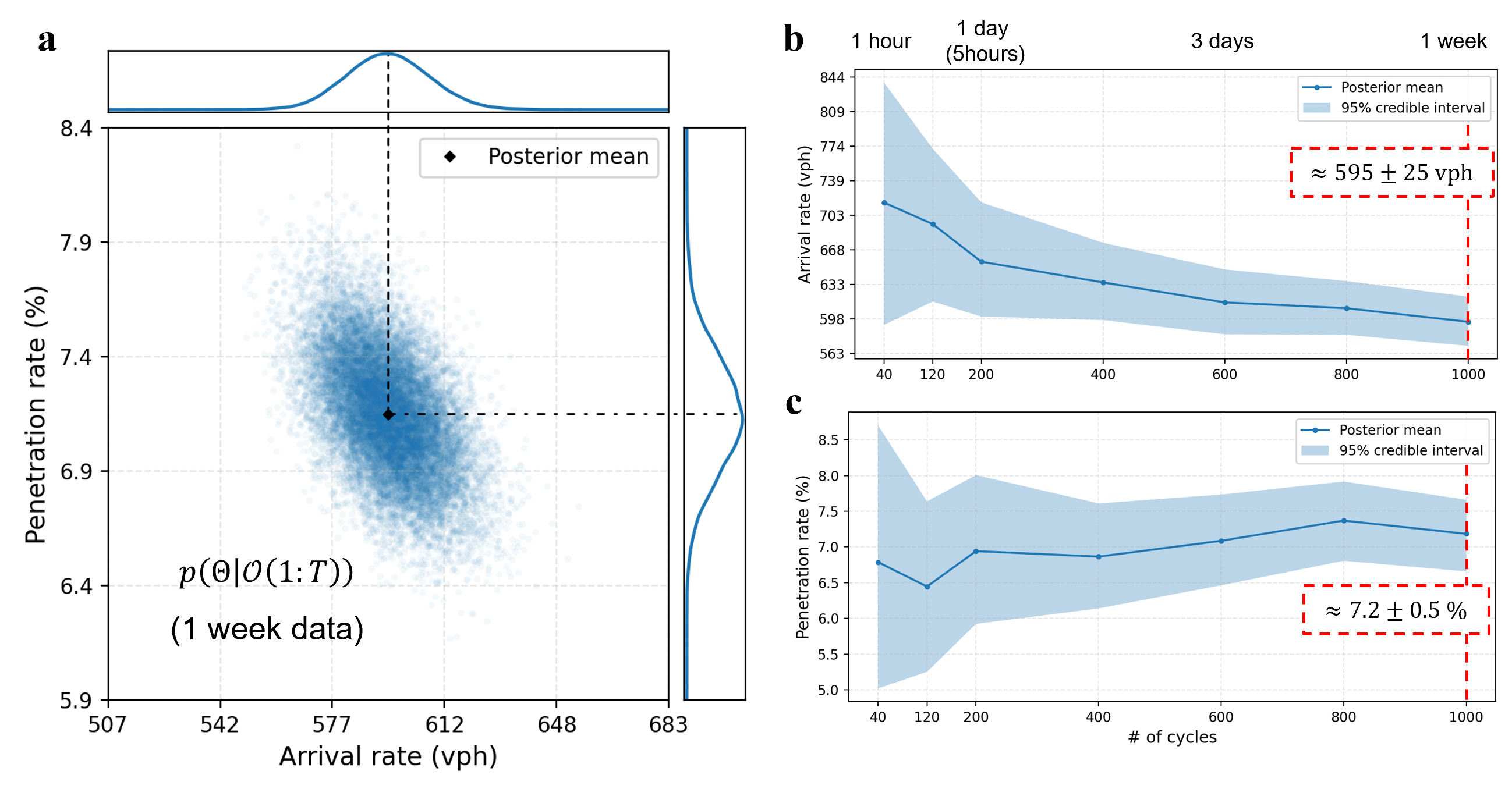}
    \caption{Distributional estimation of penetration rates and arrival rates. (a) Joint distribution of the estimated arrival rate and penetration rate using the whole week's data. (b-c) Posterior mean as well as 95\% credible interval for estimated values with different amounts of input data. }
    \label{fig:case-data-sufficiency}
\end{figure}

Figure \ref{fig:case-data-sufficiency} (a) shows the joint posterior distribution of the penetration rate $\phi$ and the average traffic volume $\mu$. The average traffic volume is calculated by:
\begin{equation}
    \mu = \sum_i \mu_i\cdot \frac{t_{i+1}-t_i}{C}.
\end{equation}
Similar to what we have observed in the simulation experiment, penetration rate $\phi$ and average traffic volume $\mu$ are negatively correlated with each other. Figure \ref{fig:case-data-sufficiency} (a) also shows the marginal distributions of both parameters. Figure \ref{fig:case-data-sufficiency} (b-c) shows the posterior mean as well as the 95\% credible interval under different amounts of input data. As expected, the estimation uncertainty decreases with the increase of given data. When $25$-hour ($1000$ cycles) data is utilized, the 95\% credible interval of the arrival rate is $[570, 620]$ vph, which is approximately $\pm 4.2\%$ of the average estimated traffic volume $595$ vph. This roughly means that, by utilizing data for a week (a total of 25-hour data, 5 hours each day), we have $95\%$ confidence that the estimation error of the average traffic volume is within $4.2\%$. 

To sum up, Figure \ref{fig:case-data-sufficiency} (b-c) clearly demonstrates how the estimation uncertainty changes with different given data sizes, which can be used as guidance with regard to the amount of data that is needed. Note that the uncertainty of the system includes both the epistemic uncertainty caused by the limited available data and the aleatoric uncertainty which refers to the intrinsic uncertainty of the arrival process. More available data only contributes to the reduction of epistemic uncertainty. Even if we have an infinite amount of data and the epistemic uncertainty goes to zero, the arrival process itself is still a stochastic process. 

% The arrival profile in Figure \ref{fig:case-arrival-est} shows the arrival probability for each time, which itself describes a random arrival process (no-homogeneous Poisson process as time interval $\Delta t\rightarrow 0$).

\begin{figure}[h!]
    \centering
    \includegraphics[width=0.65\linewidth]{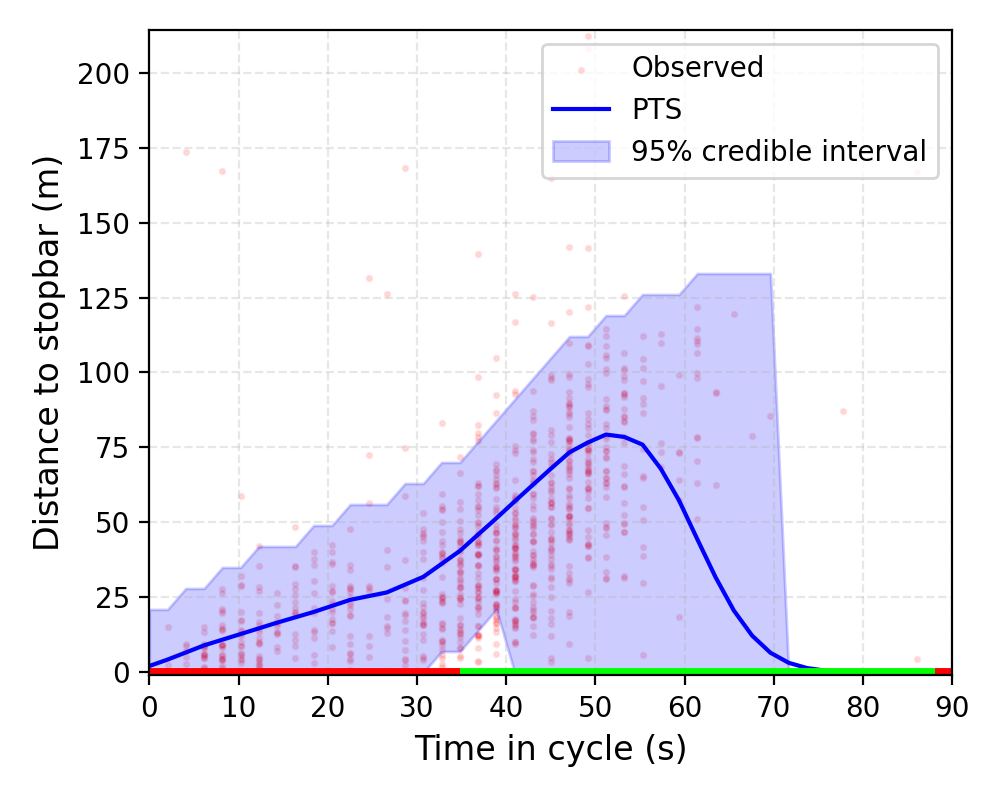}
    \caption{Estimated queue profile and observed queue scatters. }
    \label{fig:queue-scatter}
\end{figure}

Unlike the simulation studies in the previous section, we do not have the overall background traffic that can be used as the ground truth in this case study. We can only perform cross-validation for the estimation results. Figure \ref{fig:queue-scatter} shows the observed stop locations from vehicle trajectories as well as the model-estimated queue length distribution. Each red dot corresponds to one observed vehicle trajectory: the horizontal axis is the time when the trajectory starts to stop in the Newellian coordinates while the vertical axis is the distance to the stop bar. The blue line denotes the expected queue length that is calculated based on the PTS model, by taking the estimated arrival profile as input. The blue area denotes the 95\% credible interval. As shown in the figure, the model-estimated queue profile matches the observation well, which demonstrates the effectiveness of the proposed methods. Many of the outliers in the figure come from those slow-down vehicles with a short stop duration, which tend to have larger noise when we extract the observed queue length.

\section{Conclusion and discussion}  \label{sec:conclusion}

In this paper, we aim to answer one critical question: whether the available low penetration rate vehicle trajectory data is sufficient to estimate those unknown traffic states and parameters near signalized intersections. To answer this question, this paper develops a Bayesian estimation approach based on the probabilistic time-space (PTS) model. The partially observable system is formulated as a hidden Markov model (HMM). Based on this HMM formulation, the overall estimation problem is decomposed into two sub-problems: 1) traffic parameter estimation (unknown parameters in the HMM) and 2) real-time traffic state estimation (hidden states in the HMM). We develop estimation algorithms for both sub-problems, which are all established based on a single recursive algorithm. As a Bayesian approach, the proposed estimation method provides distributional estimation results and explicitly quantifies the estimation uncertainty. This paper uses simulation studies to validate the proposed method and also includes a case study using real-world trajectory data. 

% Both simulation and real-world case studies provide distributional estimation results and directly quantify the uncertainty of these estimated values. Simulation results show that the amount of required data varies in different settings of the signalized intersection. Although there is no generic answer to the data sufficiency question, a general insight implies that the vehicle delay and stop contain more useful information for the estimation of unknown traffic states and parameters. For example, at the same penetration rate, less data is needed for a movement if it is more congested (i.e., higher traffic volumes and less green split). A more straightforward conclusion is that a higher penetration rate is always preferable and beneficial which can improve the estimation accuracy. 

% In practice, we need to aggregate historical data to estimate these unknown traffic parameters by assuming that they are stationary within the same time of day (TOD). We can always choose to aggregate more data to improve the estimation accuracy despite that the stationary assumption might be less valid. On the other side, however, only recent observations are useful for the estimation of the real-time traffic states; historical data can only be utilized to provide a better prior. Consequently, the real-time traffic state estimation is largely determined by how much streaming data we have.

Other than its ability for uncertainty quantification, the proposed method has some other advantages through the utilization of a stochastic traffic flow model. First, it can fully utilize the observed data. Based on the given assumptions and simplifications, the free flow arrival time and stop locations for each trajectory compose a sufficient statistic for the estimation problem. The proposed method is able to fully utilize them through the HMM formulation. However, many previous studies did not fully utilize the observed data, including those studies that only utilized the observed stop locations at certain snapshots \citep{comert2009queue,comert2011analytical, comert2013simple,zhao2019estimation,zhao2019various,zhao2021hidden,zhao2021maximum,wong2019estimation,jia2023uncertainty}. As a result, they would significantly overestimate the uncertainty. Besides, with built-in dynamics, the proposed method can deal with over-saturation easily and naturally. Otherwise, the commonly used i.i.d. (independent and identically distributed) queue length assumption will be violated when there is a residual queue and additional efforts need to be made \citep{zhao2021hidden}. 

% Third, it can provide more complete and comprehensive estimation results. While many studies only focused on the estimation of certain parameters such as penetration rate or queue length at certain snapshots, the proposed method based on the PTS model is able to estimate the queue profile at each time and all latent variables including penetration rate and those parameters associated with vehicle arrival process.

While this paper has shed light on how to quantify the estimation uncertainty through a Bayesian approach, some efforts can be made for further improvement. First, as we briefly discussed in Section \ref{sec:observation-model}, a variety of factors would contribute to the noisy observed queue length. In this paper, we use a simple Gaussian kernel function as a simplification, which can be further improved by introducing more detailed noise models. Second, the current Assumption \ref{assump:stationary} assumes a stationary traffic demand within a certain time of day. Although it is a reasonable assumption that is frequently used in practice, it may be further relaxed. For example, a Gaussian process might be a better choice to capture the time-varying nature of the traffic demand. Third, this paper focuses on a single movement of a signalized intersection. It could be extended to a network while the main difficulty will be the curse of dimensionality as well as the resulting computational efficiency. Bayesian methods themselves are more computationally costly than point estimators. Essential decomposition techniques might be needed to simplify the problem when it comes to larger scale. We leave these for future studies.

\newpage

% CASE 1: BiBTeX used to constantly update the references
%   (while the paper is being written).
\bibliographystyle{ormsv080} % outcomment this and next line in Case 1
\bibliography{mybib} % if more than one, comma separated

\begin{thebibliography}{39}
\expandafter\ifx\csname natexlab\endcsname\relax\def\natexlab#1{#1}\fi
\expandafter\ifx\csname url\endcsname\relax
  \def\url#1{{\tt #1}}\fi
\expandafter\ifx\csname urlprefix\endcsname\relax\def\urlprefix{URL }\fi
\expandafter\ifx\csname urlstyle\endcsname\relax
  \expandafter\ifx\csname doi\endcsname\relax
  \def\doi#1{doi:\discretionary{}{}{}#1}\fi \else
  \expandafter\ifx\csname doi\endcsname\relax
  \def\doi{doi:\discretionary{}{}{}\begingroup \urlstyle{rm}\Url}\fi \fi

\bibitem[{Abril-Pla et~al.(2023)Abril-Pla, Andreani, Carroll, Dong, Fonnesbeck, Kochurov, Kumar, Lao, Luhmann, Martin et~al.}]{abril2023pymc}
Abril-Pla, Oriol, Virgile Andreani, Colin Carroll, Larry Dong, Christopher~J Fonnesbeck, Maxim Kochurov, Ravin Kumar, Junpeng Lao, Christian~C Luhmann, Osvaldo~A Martin, et~al. 2023.
\newblock Pymc: a modern, and comprehensive probabilistic programming framework in python.
\newblock {\it PeerJ Computer Science\/} {\bf 9} e1516.

\bibitem[{Axer and Friedrich(2017)}]{axer2017signal}
Axer, Steffen, Bernhard Friedrich. 2017.
\newblock Signal timing estimation based on low frequency floating car data.
\newblock {\it Transportation research procedia\/} {\bf 25} 1645--1661.

\bibitem[{Ban et~al.(2011)Ban, Hao, and Sun}]{ban2011real}
Ban, Xuegang~Jeff, Peng Hao, Zhanbo Sun. 2011.
\newblock Real time queue length estimation for signalized intersections using travel times from mobile sensors.
\newblock {\it Transportation Research Part C: Emerging Technologies\/} {\bf 19}(6) 1133--1156.

\bibitem[{Cheng et~al.(2012)Cheng, Qin, Jin, and Ran}]{cheng2012exploratory}
Cheng, Yang, Xiao Qin, Jing Jin, Bin Ran. 2012.
\newblock An exploratory shockwave approach to estimating queue length using probe trajectories.
\newblock {\it Journal of intelligent transportation systems\/} {\bf 16}(1) 12--23.

\bibitem[{Comert(2013)}]{comert2013simple}
Comert, Gurcan. 2013.
\newblock Simple analytical models for estimating the queue lengths from probe vehicles at traffic signals.
\newblock {\it Transportation Research Part B: Methodological\/} {\bf 55} 59--74.

\bibitem[{Comert and Cetin(2009)}]{comert2009queue}
Comert, Gurcan, Mecit Cetin. 2009.
\newblock Queue length estimation from probe vehicle location and the impacts of sample size.
\newblock {\it European Journal of Operational Research\/} {\bf 197}(1) 196--202.

\bibitem[{Comert and Cetin(2011)}]{comert2011analytical}
Comert, Gurcan, Mecit Cetin. 2011.
\newblock Analytical evaluation of the error in queue length estimation at traffic signals from probe vehicle data.
\newblock {\it IEEE Transactions on Intelligent Transportation Systems\/} {\bf 12}(2) 563--573.

\bibitem[{Doucet et~al.(2009)Doucet, Johansen et~al.}]{doucet2009tutorial}
Doucet, Arnaud, Adam~M Johansen, et~al. 2009.
\newblock A tutorial on particle filtering and smoothing: Fifteen years later.
\newblock {\it Handbook of nonlinear filtering\/} {\bf 12}(656-704) 3.

\bibitem[{Du et~al.(2019)Du, Yan, Zhu, and Sun}]{du2019signal}
Du, Zelong, Xintao Yan, Jinqing Zhu, Weili Sun. 2019.
\newblock Signal timing parameters estimation for intersections using floating car data.
\newblock {\it Transportation research record\/} {\bf 2673}(6) 189--201.

\bibitem[{Gelman et~al.(2013)Gelman, Carlin, Stern, Dunson, Vehtari, and Rubin}]{gelman2013bayesian}
Gelman, Andrew, John~B Carlin, Hal~S Stern, David~B Dunson, Aki Vehtari, Donald~B Rubin. 2013.
\newblock {\it Bayesian data analysis\/}.
\newblock CRC press.

\bibitem[{Guo et~al.(2019)Guo, Li, and Ban}]{guo2019urban}
Guo, Qiangqiang, Li~Li, Xuegang~Jeff Ban. 2019.
\newblock Urban traffic signal control with connected and automated vehicles: A survey.
\newblock {\it Transportation research part C: emerging technologies\/} {\bf 101} 313--334.

\bibitem[{Hoffman et~al.(2014)Hoffman, Gelman et~al.}]{hoffman2014no}
Hoffman, Matthew~D, Andrew Gelman, et~al. 2014.
\newblock The no-u-turn sampler: adaptively setting path lengths in hamiltonian monte carlo.
\newblock {\it J. Mach. Learn. Res.\/} {\bf 15}(1) 1593--1623.

\bibitem[{Jia et~al.(2023)Jia, Wong, and Wong}]{jia2023uncertainty}
Jia, Shaocheng, SC~Wong, Wai Wong. 2023.
\newblock Uncertainty estimation of connected vehicle penetration rate.
\newblock {\it Transportation Science\/} .

\bibitem[{Krajzewicz et~al.(2012)Krajzewicz, Erdmann, Behrisch, and Bieker}]{krajzewicz2012recent}
Krajzewicz, Daniel, Jakob Erdmann, Michael Behrisch, Laura Bieker. 2012.
\newblock Recent development and applications of sumo-simulation of urban mobility.
\newblock {\it International journal on advances in systems and measurements\/} {\bf 5}(3\&4).

\bibitem[{Li et~al.(2017)Li, Tang, Yao, and Li}]{li2017real}
Li, Fuliang, Keshuang Tang, Jiarong Yao, Keping Li. 2017.
\newblock Real-time queue length estimation for signalized intersections using vehicle trajectory data.
\newblock {\it Transportation Research Record\/} {\bf 2623}(1) 49--59.

\bibitem[{Light and Whitham(1955)}]{light1955kinematic}
Light, MJ, B~Whitham. 1955.
\newblock On kinematic waves. i: Flow movement in long rivers; ii: A theory of traffic flow on long crowed roads [c].
\newblock {\it Proceedings of Royal Society A\/} (229) 281--345.

\bibitem[{Lloret-Batlle and Zheng(2023)}]{lloret2023jam}
Lloret-Batlle, Roger, Jianfeng Zheng. 2023.
\newblock Jam density and stopbar location estimation with trajectory data at signalized intersections.
\newblock {\it Transportation Research Part B: Methodological\/} {\bf 173} 162--175.

\bibitem[{MacKay(2003)}]{mackay2003information}
MacKay, David~JC. 2003.
\newblock {\it Information theory, inference and learning algorithms\/}.
\newblock Cambridge university press.

\bibitem[{Maripini et~al.(2023)Maripini, Khadhir, and Vanajakshi}]{maripini2023traffic}
Maripini, Himabindu, Abdhul Khadhir, Lelitha Vanajakshi. 2023.
\newblock Traffic state estimation near signalized intersections.
\newblock {\it Journal of Transportation Engineering, Part A: Systems\/} {\bf 149}(5) 03123002.

\bibitem[{Neal and Hinton(1998)}]{neal1998view}
Neal, Radford~M, Geoffrey~E Hinton. 1998.
\newblock A view of the em algorithm that justifies incremental, sparse, and other variants.
\newblock {\it Learning in graphical models\/}. Springer, 355--368.

\bibitem[{Newell(2002)}]{newell2002simplified}
Newell, Gordon~Frank. 2002.
\newblock A simplified car-following theory: a lower order model.
\newblock {\it Transportation Research Part B: Methodological\/} {\bf 36}(3) 195--205.

\bibitem[{Ramezani and Geroliminis(2015)}]{ramezani2015queue}
Ramezani, Mohsen, Nikolas Geroliminis. 2015.
\newblock Queue profile estimation in congested urban networks with probe data.
\newblock {\it Computer-Aided Civil and Infrastructure Engineering\/} {\bf 30}(6) 414--432.

\bibitem[{Richards(1956)}]{richards1956shock}
Richards, Paul~I. 1956.
\newblock Shock waves on the highway.
\newblock {\it Operations research\/} {\bf 4}(1) 42--51.

\bibitem[{Saldivar-Carranza et~al.(2021)Saldivar-Carranza, Li, Mathew, Hunter, Sturdevant, and Bullock}]{saldivar2021deriving}
Saldivar-Carranza, Enrique, Howell Li, Jijo Mathew, Margaret Hunter, James Sturdevant, Darcy~M Bullock. 2021.
\newblock Deriving operational traffic signal performance measures from vehicle trajectory data.
\newblock {\it Transportation research record\/} {\bf 2675}(9) 1250--1264.

\bibitem[{Seo et~al.(2017)Seo, Bayen, Kusakabe, and Asakura}]{seo2017traffic}
Seo, Toru, Alexandre~M Bayen, Takahiko Kusakabe, Yasuo Asakura. 2017.
\newblock Traffic state estimation on highway: A comprehensive survey.
\newblock {\it Annual reviews in control\/} {\bf 43} 128--151.

\bibitem[{Sun and Ban(2013)}]{sun2013vehicle}
Sun, Zhanbo, Xuegang~Jeff Ban. 2013.
\newblock Vehicle trajectory reconstruction for signalized intersections using mobile traffic sensors.
\newblock {\it Transportation Research Part C: Emerging Technologies\/} {\bf 36} 268--283.

\bibitem[{Waddell et~al.(2020)Waddell, Remias, Kirsch, and Young}]{waddell2020scalable}
Waddell, Jonathan~M, Stephen~M Remias, Jenna~N Kirsch, Stanley~E Young. 2020.
\newblock Scalable and actionable performance measures for traffic signal systems using probe vehicle trajectory data.
\newblock {\it Transportation Research Record\/} {\bf 2674}(11) 304--316.

\bibitem[{Wang et~al.(2024)Wang, Jerome, Wang, Zhang, Shen, Kumar, Bai, Krajewski, Deneau, Jawad et~al.}]{wang2024osaas}
Wang, Xingmin, Zachary Jerome, Zihao Wang, Chenhao Zhang, Shengyin Shen, Vivek~Vijaya Kumar, Fan Bai, Paul Krajewski, Danielle Deneau, Ahmad Jawad, et~al. 2024.
\newblock Traffic light optimization with low penetration rate vehicle trajectory data.
\newblock {\it Nature Communications\/} {\bf 15}(1) 1306.

\bibitem[{Wang et~al.(2023)Wang, Jerome, Zhang, Shen, Kumar, and Liu}]{wang2023trajectory}
Wang, Xingmin, Zachary Jerome, Chenhao Zhang, Shengyin Shen, Vivek~Vijaya Kumar, Henry~X Liu. 2023.
\newblock Trajectory data processing and mobility performance evaluation for urban traffic networks.
\newblock {\it Transportation Research Record\/} {\bf 2677}(3) 355--370.

\bibitem[{Wang et~al.(2022)Wang, Zhao, Yu, Hu, Zheng, Hua, Zhang, Hu, and Guo}]{wang2022real}
Wang, Yibing, Mingming Zhao, Xianghua Yu, Yonghui Hu, Pengjun Zheng, Wei Hua, Lihui Zhang, Simon Hu, Jingqiu Guo. 2022.
\newblock Real-time joint traffic state and model parameter estimation on freeways with fixed sensors and connected vehicles: State-of-the-art overview, methods, and case studies.
\newblock {\it Transportation Research Part C: Emerging Technologies\/} {\bf 134} 103444.

\bibitem[{Welch et~al.(1995)Welch, Bishop et~al.}]{welch1995introduction}
Welch, Greg, Gary Bishop, et~al. 1995.
\newblock An introduction to the kalman filter .

\bibitem[{Wong et~al.(2019)Wong, Shen, Zhao, and Liu}]{wong2019estimation}
Wong, Wai, Shengyin Shen, Yan Zhao, Henry~X Liu. 2019.
\newblock On the estimation of connected vehicle penetration rate based on single-source connected vehicle data.
\newblock {\it Transportation Research Part B: Methodological\/} {\bf 126} 169--191.

\bibitem[{Xing et~al.(2022)Xing, Wu, Cheng, and Liu}]{xing2022traffic}
Xing, Jiping, Wei Wu, Qixiu Cheng, Ronghui Liu. 2022.
\newblock Traffic state estimation of urban road networks by multi-source data fusion: Review and new insights.
\newblock {\it Physica A: Statistical Mechanics and its Applications\/} {\bf 595} 127079.

\bibitem[{Yao et~al.(2019)Yao, Li, Tang, and Jian}]{yao2019sampled}
Yao, Jiarong, Fuliang Li, Keshuang Tang, Sun Jian. 2019.
\newblock Sampled trajectory data-driven method of cycle-based volume estimation for signalized intersections by hybridizing shockwave theory and probability distribution.
\newblock {\it IEEE Transactions on Intelligent Transportation Systems\/} {\bf 21}(6) 2615--2627.

\bibitem[{Zhao et~al.(2021{\natexlab{a}})Zhao, Shen, and Liu}]{zhao2021hidden}
Zhao, Yan, Shengyin Shen, Henry~X Liu. 2021{\natexlab{a}}.
\newblock A hidden markov model for the estimation of correlated queues in probe vehicle environments.
\newblock {\it Transportation Research Part C: Emerging Technologies\/} {\bf 128} 103128.

\bibitem[{Zhao et~al.(2021{\natexlab{b}})Zhao, Wong, Zheng, and Liu}]{zhao2021maximum}
Zhao, Yan, Wai Wong, Jianfeng Zheng, Henry~X Liu. 2021{\natexlab{b}}.
\newblock Maximum likelihood estimation of probe vehicle penetration rates and queue length distributions from probe vehicle data.
\newblock {\it IEEE Transactions on Intelligent Transportation Systems\/} {\bf 23}(7) 7628--7636.

\bibitem[{Zhao et~al.(2019{\natexlab{a}})Zhao, Zheng, Wong, Wang, Meng, and Liu}]{zhao2019estimation}
Zhao, Yan, Jianfeng Zheng, Wai Wong, Xingmin Wang, Yuan Meng, Henry~X Liu. 2019{\natexlab{a}}.
\newblock Estimation of queue lengths, probe vehicle penetration rates, and traffic volumes at signalized intersections using probe vehicle trajectories.
\newblock {\it Transportation Research Record\/} {\bf 2673}(11) 660--670.

\bibitem[{Zhao et~al.(2019{\natexlab{b}})Zhao, Zheng, Wong, Wang, Meng, and Liu}]{zhao2019various}
Zhao, Yan, Jianfeng Zheng, Wai Wong, Xingmin Wang, Yuan Meng, Henry~X Liu. 2019{\natexlab{b}}.
\newblock Various methods for queue length and traffic volume estimation using probe vehicle trajectories.
\newblock {\it Transportation Research Part C: Emerging Technologies\/} {\bf 107} 70--91.

\bibitem[{Zheng and Liu(2017)}]{zheng2017estimating}
Zheng, Jianfeng, Henry~X Liu. 2017.
\newblock Estimating traffic volumes for signalized intersections using connected vehicle data.
\newblock {\it Transportation Research Part C: Emerging Technologies\/} {\bf 79} 347--362.

\end{thebibliography}

% CASE 2: BiBTeX used to generate mypaper.bbl (to be further fine tuned)
%\input{mypaper.bbl} % outcomment this line in Case 2

%If you don't use BiBTex, you can manually itemize references as shown below.

\AUTHORBIO{
		% \textbf{Jianzhe Zhen} is a final year PhD student at the Department of Econometrics and Operations Research at Tilburg University. His research is focused on robust optimization. He won the Student Best Paper Prize for this paper at the Computational Management Science 2017 conference in Bergamo. \\
		
		% \textbf{Dick den Hertog} is professor of Business Analytics \& Operations Research at Tilburg University.
		% His research interests cover various fields in prescriptive analytics, in particular linear and nonlinear optimization. In recent years his main focus has been on robust optimization and simulation-based optimization. He is also active in applying the theory in real-life applications. In particular, he is interested in applications that contribute to a better society. For many years he has been involved in research to optimize water safety, he is doing research to develop better optimization models and techniques for cancer treatment, and recently he got involved in research to optimize the food supply chain for World Food Programme. In 2000 he received the EURO Best Applied Paper Award, together with Peter Stehouwer (CQM). In 2013 he was a member of the team that received the INFORMS Franz Edelman Award.\\
		
		% \textbf{Melvyn Sim} is a professor at the Department of Analytics \& Operations, NUS Business school. His research interests fall broadly under the categories of decision making and optimization under uncertainty with applications ranging from finance, supply chain management, healthcare to engineered systems.
}		
		
		%----------------------------------------------------------------------------------------
		
	\end{document}